\documentclass[notitlepage,aps,12pt,tightenlines,nofootinbib]{revtex4}
\pdfoutput=1

\usepackage{amsmath}
\usepackage{amssymb,amsfonts}
\usepackage{graphicx}
\usepackage{subfigure}

\graphicspath{{figs/}{figs_lo/}}

\newcommand{\note}[1]{}

\newcommand{\mathnotation}[2]{\newcommand{#1}{\ensuremath{#2}}}
\newcommand{\ie}{\textit{i.e.}}
\DeclareMathOperator{\JSR}{JSR}

\mathnotation{\ldef}{\mathrel{\raisebox{.069ex}{:}\!\!=}}% Left define
\mathnotation{\nn}{n}				% Number of rods/particles
\mathnotation{\Br}{B}				% Braid group
\mathnotation{\asigma}{\Sigma}			% Generators on the annulus

\begin{document}

\title{Topological Optimisation of Rod-Stirring Devices}
\author{Matthew D. Finn}
\email{matthew.finn@adelaide.edu.au}
\affiliation{School of Mathematical Sciences, University of Adelaide,
  Adelaide SA 5005, Australia}

\author{Jean-Luc Thiffeault}
\email{jeanluc@math.wisc.edu}
\affiliation{Department of Mathematics, University
  of Wisconsin, Madison, WI 53706, USA}
\date{\today}

\begin{abstract}
  There are many industrial situations where rods are used to stir a
  fluid, or where rods repeatedly stretch a material such as bread
  dough or taffy.  The goal in these applications is to stretch either
  material lines (in a fluid) or the material itself (for dough or
  taffy) as rapidly as possible.  The growth rate of material lines is
  conveniently given by the topological entropy of the rod motion.  We
  discuss the problem of optimising such rod devices from a
  topological viewpoint.  We express rod motions in terms of
  generators of the braid group, and assign a cost based on the minimum
  number of generators needed to write the braid.  We show that for
  one cost function---the topological entropy per generator---the
  optimal growth rate is the logarithm of the golden ratio.  For a
  more realistic cost function, involving the topological entropy per
  operation where rods are allowed to move together, the optimal
  growth rate is the logarithm of the silver ratio, $1+\sqrt{2}$.  We
  show how to construct devices that realise this optimal growth,
  which we call \emph{silver mixers}.
\end{abstract}


\maketitle

\section{Introduction}
\label{sec:intro}

A rod-stirring device, where a number of rods are moved around in a
fluid, is the most natural and intuitive method of stirring.  The
number of rods, their shape, and the nature of their motion constitute
a stirring protocol.  For example, moving a spoon in a figure-eight
pattern in a mixture is a simple way of blending the ingredients of a
cake.  But beyond applications in the kitchen, such stirring methods
are widely used in industry.  For instance, when glass is
manufactured, it is in a molten state.  It is inhomogeneous both in
temperature and composition.  These inhomogeneities are undesirable:
the human eye can detect very small variations in density, and modern
glass must be of the highest quality.  One way to remove these
inhomogeneities is to stir the molten glass before it cools.

\begin{figure}
\begin{center}
\subfigure[]{
\includegraphics[width=.315\textwidth]{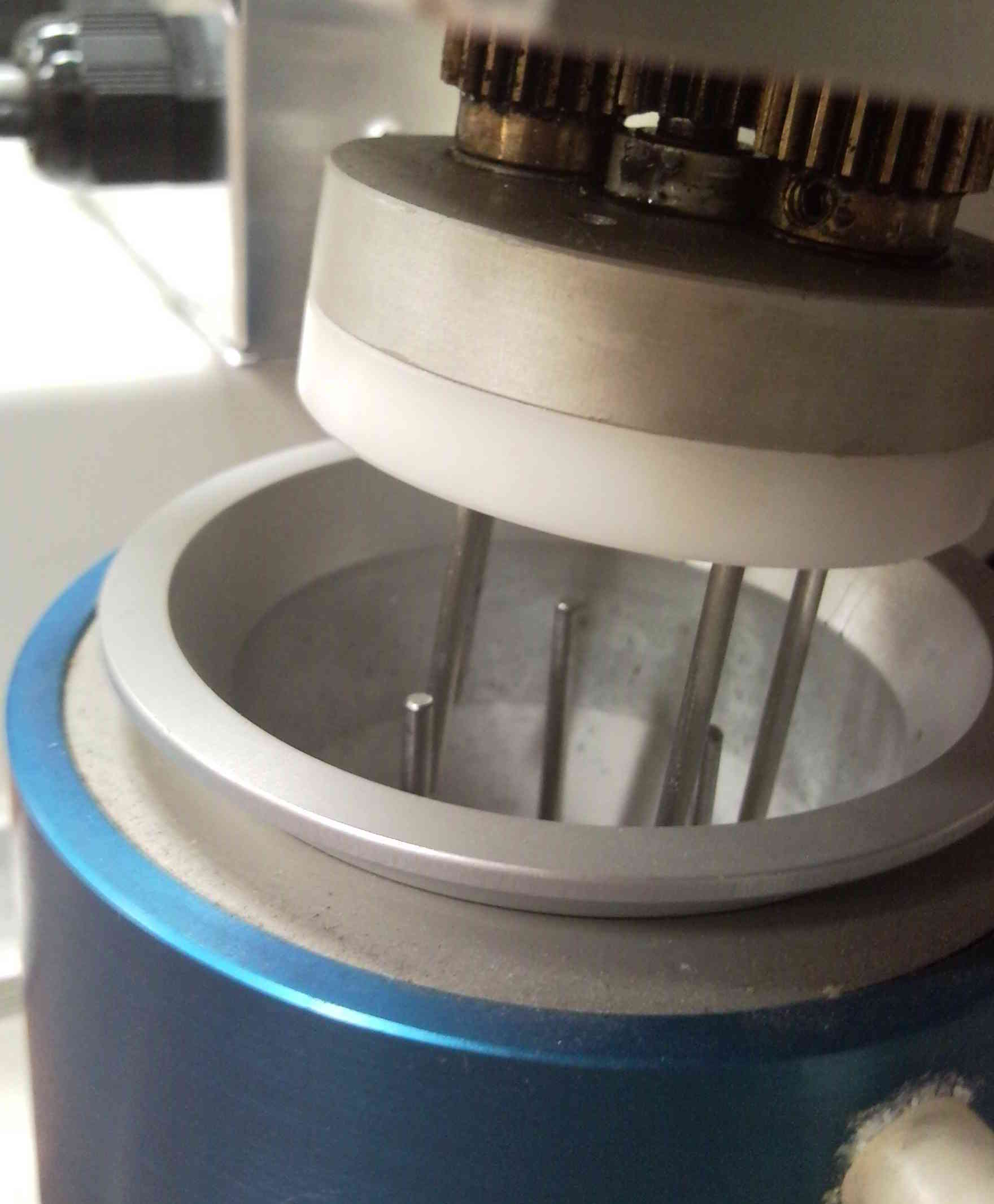}
\label{fig:pinmixer6}
}\hspace{.25in}%
\subfigure[]{
\begin{minipage}{.2\textwidth}
\vspace{-2.2in}
\includegraphics[width=\textwidth]{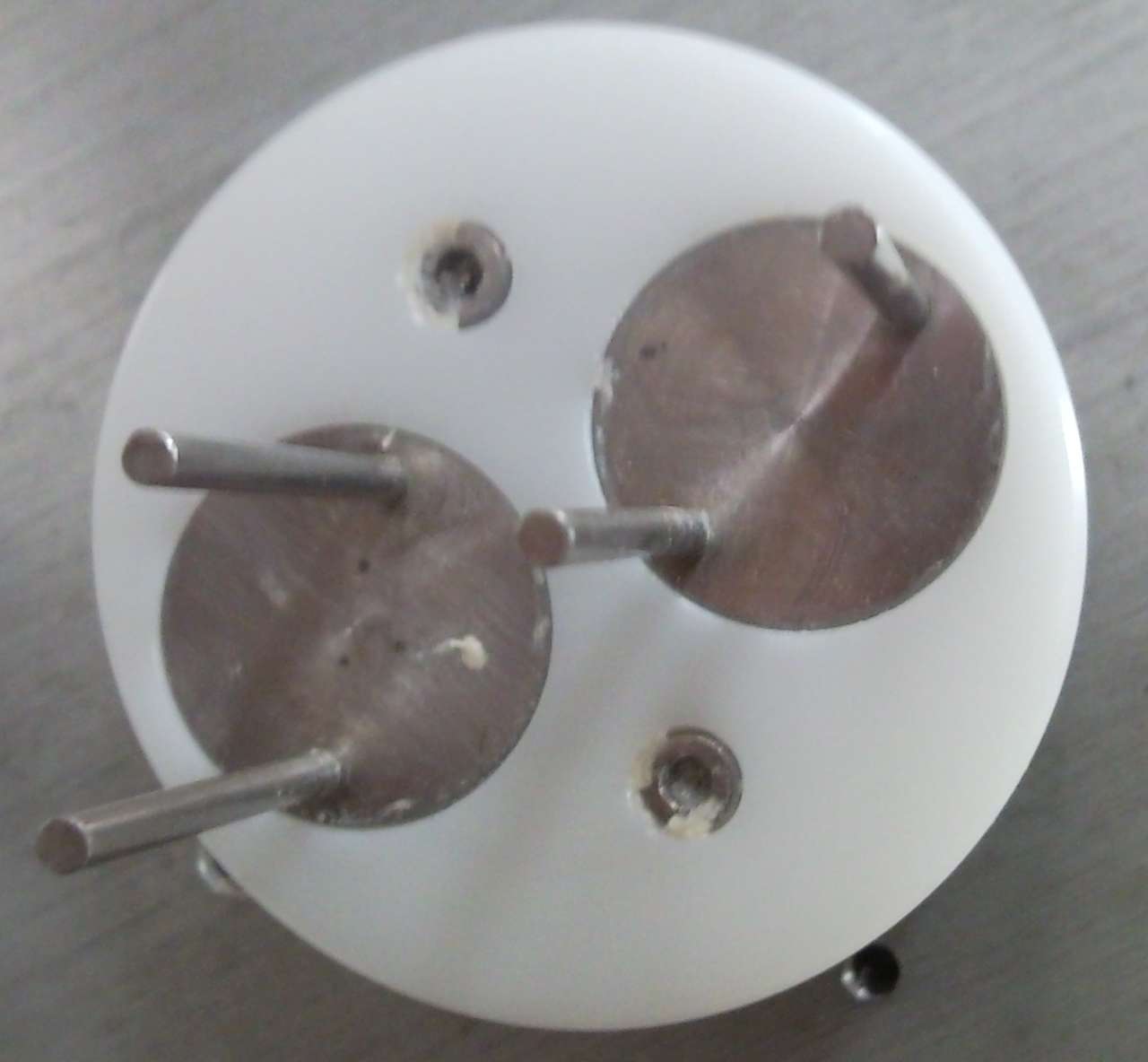}
\includegraphics[width=\textwidth]{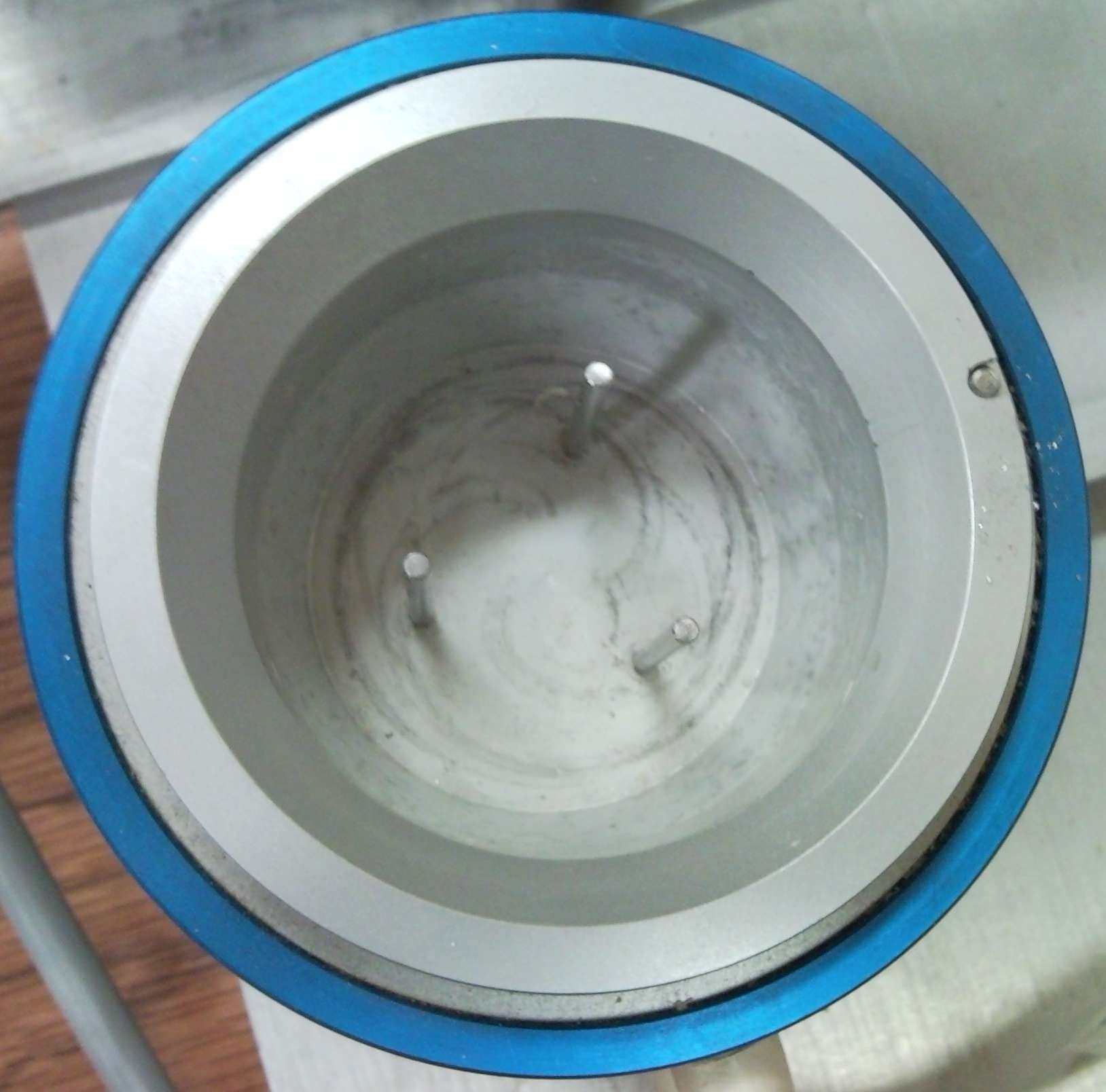}
\label{fig:pinmixer34}
\end{minipage}
}
\end{center}
\caption{(a) The mixograph, a model planetary pin mixer for bread
  dough. (b) Top section with four moving rods (above), and bottom section
  with three fixed rods (below).  See
  \citet{Connelly2008}. (Courtesy of the Department of Food Science,
  University of Wisconsin.)}
\label{fig:pinmixer}
\end{figure}

\begin{figure}
\begin{center}
\subfigure[]{
\begin{minipage}{.425\textwidth}
\includegraphics[width=\textwidth]{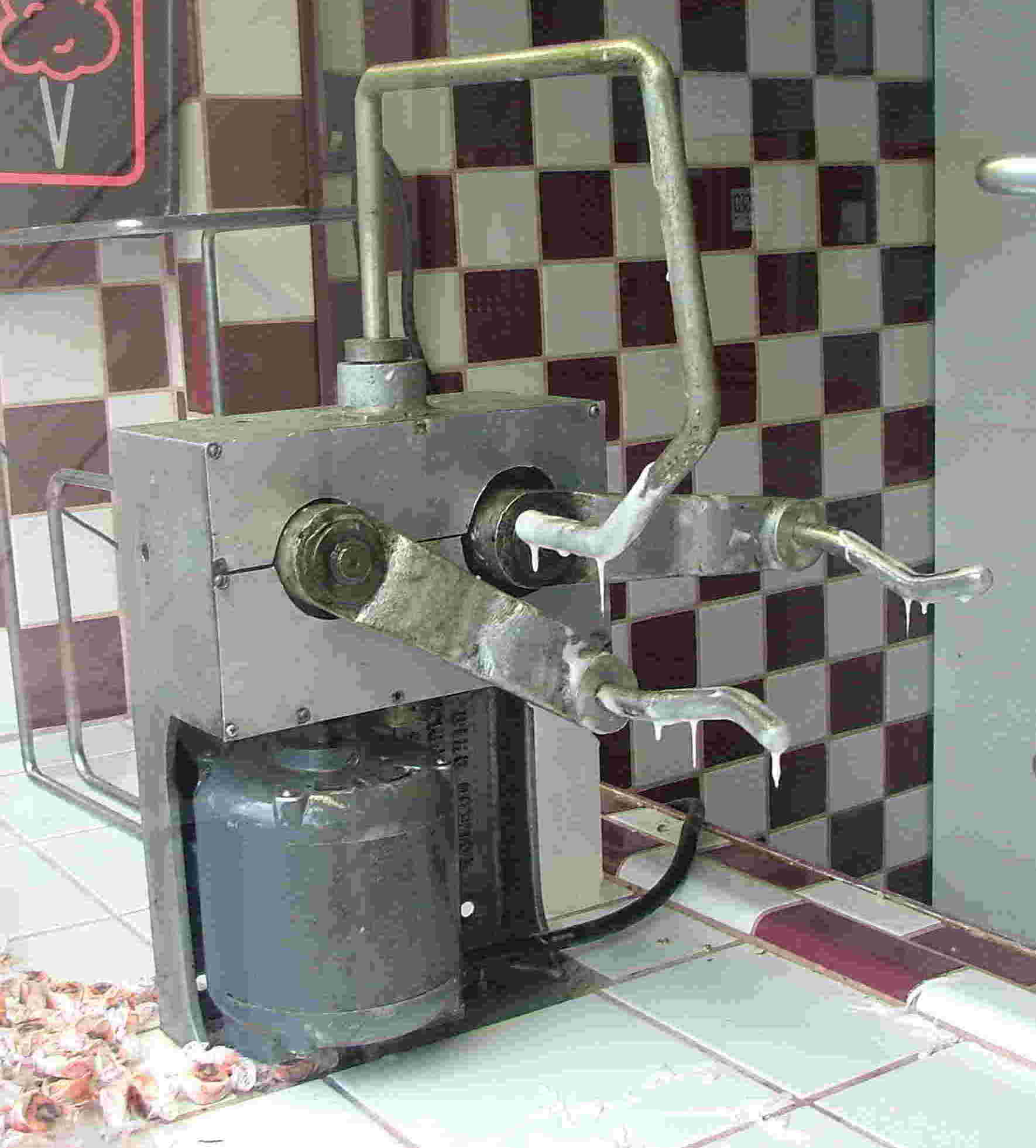}

\vspace{.75em}
\end{minipage}
%\hspace{.75em}
\label{fig:taffya}
}\hspace{.01\textwidth}%
\subfigure[]{
\begin{minipage}{.4\textwidth}
% Full period: about 187 frames.
\includegraphics[width=.3\textwidth]{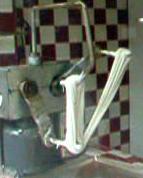}
\hspace*{-.09em}
\includegraphics[width=.3\textwidth]{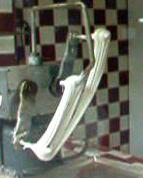}
\hspace*{-.09em}
\includegraphics[width=.3\textwidth]{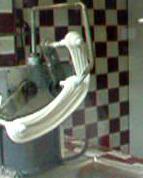}\\[4pt]

\includegraphics[width=.3\textwidth]{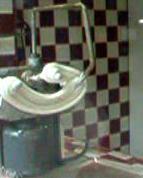}
\hspace*{-.09em}
\includegraphics[width=.3\textwidth]{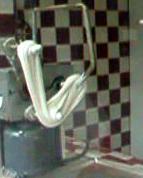}
\hspace*{-.09em}
\includegraphics[width=.3\textwidth]{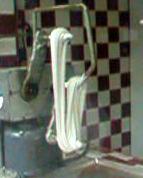}\\[4pt]

\includegraphics[width=.3\textwidth]{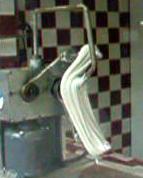}
\hspace*{-.09em}
\includegraphics[width=.3\textwidth]{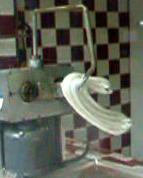}
\hspace*{-.09em}
\includegraphics[width=.3\textwidth]{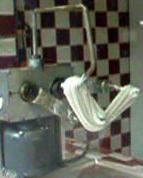}

\vspace{.75em}
\end{minipage}
\label{fig:taffyb}
}
\end{center}
\caption{(a) A taffy-pulling device. (b) Snapshots over a full period
  of operation, with taffy.}
\label{fig:taffy}
\end{figure}

There, problems begin: you cannot stir molten glass the way you would
a cup of coffee---it is simply too viscous.  Hence, the stirring rods
are limited in their speed and path, and it is important to know what
paths are optimal in order to improve the performance of a mixing device.
This is where mathematics can help, and is the topic of this paper: we
introduce a topological approach to the optimisation of rod-stirring
devices.  The topological approach was pioneered for fluids
by~\citet{Boyland2000} and developed by many others~\cite{Boyland1994,
  Boyland2003, MattFinn2003, MattFinn2003b, Vikhansky2004,
  Thiffeault2005, Gouillart2006, MattFinn2006, Moussafir2006,
  Thiffeault2006, MattFinn2007, Stremler2007, Binder2008,
  Thiffeault2008b, Thiffeault2009, Thiffeault2010}.  Of course, glass
mixing is only a representative application, and the work presented
here applies to many other important forms of mixing, such as
industrial dough production which uses a type of device called a pin
mixer~\cite{Connelly2008} (see Fig.~\ref{fig:pinmixer}).  In many
cases, as in taffy-making (see Fig.~\ref{fig:taffy}), there is not
necessarily an underlying `fluid' filling the space, but the
optimisation techniques we discuss apply directly and so we use the
same language to describe these devices.

In all physical problems, modelling begins by making simplifying
assumptions.  For glass mixing and for many other situations, the fluid
motion is predominantly two-dimensional.  This is due in part to the
thinness of the glass layer, but also to its thermal stratification.
Hence, it is suitable and highly desirable to model the system as an
idealised two-dimensional fluid.  Mathematically, this means that the
domain of fluid motion is a two-dimensional space---a surface.  More
precisely, it is a punctured surface with an outer boundary, since the
stirring creates obstacles in the space around which the fluid must
flow.  Hence, if we are only interested in a topological
characterisation of the mixing region, we can consider our simplified
mixing device to be the punctured disc, where the movable punctures
are the stirring rods.  By topological characterisation, we mean that
we do not care about the specific size of features, only about their
global impact on the space.

The fluid motion induced by any periodic motion of the rods (or
punctures) defines a mapping, that is, a rule for describing how small
parcels of fluid are dragged by the rods at each stirring cycle.  This
mapping is normally obtained by solving the relevant fluid equations:
the Stokes equations for highly-viscous fluids such as molten glass,
or more complicated equations when stirring a polymeric fluid.  But
the beauty of a topological approach is that we can deduce much 
about the range of possible outcomes of rod motion without ever
specifying governing equations of motion.  From this point of view, it
suffices that the fluid is an idealised continuum, an assumption that
holds for essentially all fluids (and even granular materials in some
limits).

Hence, the mathematical setting for our problem is the space of all
possible mappings of the punctured disc that arise from rod
motions. This is a gigantic space, and it contains as special cases
all possible fluid motions.  The key to making sense of this space is
that there is a theorem---the Thurston--Nielsen classification
theorem~\cite{Fathi1979,Thurston1988}---that tells us that the members
of that space (the mappings induced by rod motions) can be grouped
into three (and only three!) categories.

The first category is called finite-order.  If we move the rods around
following this type of motion, and return them to their original
position, then in a well-defined sense material lines in the fluid
have not become `entwined' on the rods.  (These mappings are isotopic
to the identity.)  Since our ultimate goal is to mix the fluid by
stirring, this is terrible.  It means that we are wasting our precious
time and energy with this particular rod motion.  We shall therefore say nothing
more about this category.

The second category of mappings is called pseudo-Anosov, and it is
much more interesting.  A pseudo-Anosov mapping leads to a complex
intertwining of material lines in the fluid around the rods.  In fact,
a typical material line is forced to grow exponentially under repeated
stirring.  The pseudo-Anosov mappings are closely related to chaotic
dynamical systems.  They are the best type to stir with, and for
appropriate rod motions they can be very effective.  This was
demonstrated by~\citet{Boyland2000}, who constructed a rod-stirring
device with a pseudo-Anosov rod motion.  (A mechanically more
straightforward implementation was given by~\citet{Binder2008}.)  Their
design is, however, far from optimal, and we will improve upon it in
this paper.  Note that to have a pseudo-Anosov mapping we need at
least three rods: it is not possible to have topological complexity
with only one or two rods~\cite{Boyland2000}, unless artificial
`punctures' are made in the space~\cite{Gouillart2006}.  This helps
explain why the taffy puller in Fig.~\ref{fig:taffy} has three rods.

The third category of mappings is called reducible.  Essentially,
mappings in that category break up the fluid domain into separate
regions, where each region belongs to the first or second categories
above. Mappings in this category are likely to have barriers to mixing,
which is undesirable. We shall not discuss maps of this type any further.

Our aim in this paper is to select rod motions that belong to the
second category---pseudo-Anosov---and are particularly good at mixing.
To accomplish this, we first need a convenient manner of describing
rod motions topologically.  We show how to do this in terms of braids
in Section~\ref{sec:braids}.  Next, we need to specify our measure of
`good mixing,' the topological entropy: this is the minimum complexity
imparted on the fluid trajectories based on the rod motions (see
Section~\ref{sec:entropy}).  We also need to ascribe a cost to the rod
motions, and we present two possible choices in
Section~\ref{sec:optgen}. One of these measures of cost
we argue the more physically relevant.
In Sections~\ref{sec:annulus} and~\ref{sec:silvermix} we
describe a rod-stirring device that incorporates our optimality
criterion, and show results from simulations and experiments.  We
offer some concluding remarks in Section~\ref{sec:discussion}.


\section{Rod Motions as Braids}
\label{sec:braids}

In the introduction, we saw that the mapping describing the fluid
motion belongs to one of three categories: finite-order,
pseudo-Anosov, or reducible.  Only the second is important for our
purposes.  But even within the category of pseudo-Anosov mappings,
there are many \emph{equivalence classes} of mappings.  This means
that there are mappings that are fundamentally different, in that they
cannot be obtained from one another by a continuous transformation.
This concept is called \emph{isotopy} of mappings, and the set of all
mappings isotopic to a given mapping is called its \emph{isotopy
  class}.  For a more detailed discussion see~\cite{Boyland1994,
  Boyland2000, Boyland2003, Gouillart2006, Thiffeault2005,
  Thiffeault2006, Thiffeault2008b, Thiffeault2010}.

\begin{figure}
  \begin{center}
\subfigure[]{
    \includegraphics[height=.3\textheight]{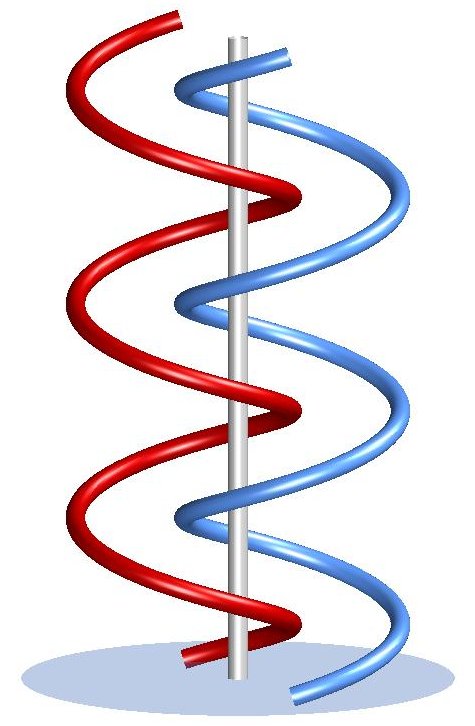}
    \label{fig:taffybraid}
}\hspace{.1\textwidth}%
\subfigure[]{
    \includegraphics[height=.25\textheight]{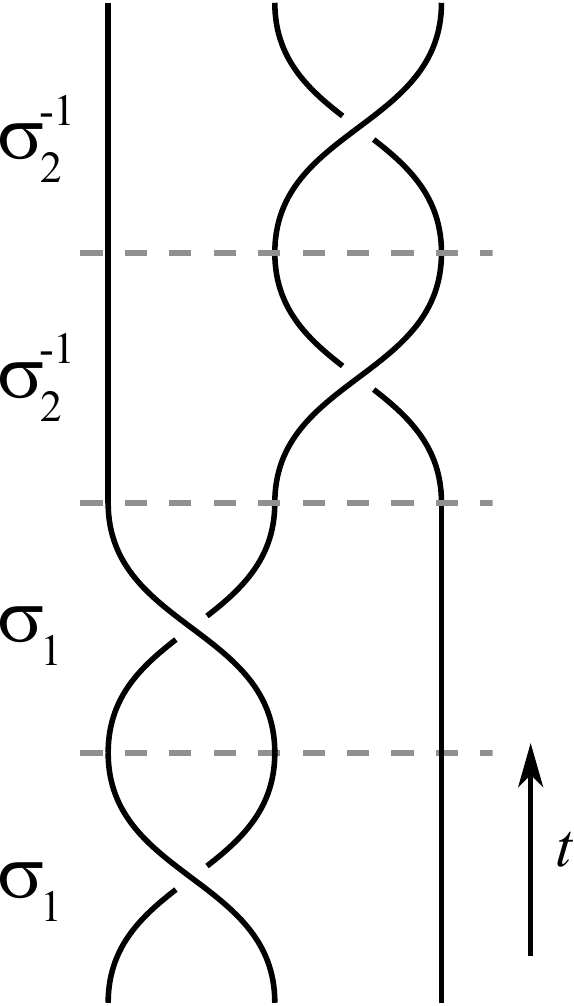}
    \label{fig:taffybraid_diagram}
}
\end{center}
\caption{(a) The trajectories of the three rods of the taffy puller
  plotted in a space-time diagram, with time flowing from bottom to
  top.  This defines a braid on~$\nn=3$
  strands, which in terms of braid group generators is
  written~$\sigma_1^2\sigma_{2}^{-2}$, read from left to right (three periods are shown).
  (b) The diagram for the braid~$\sigma_1^2\sigma_{2}^{-2}$, which has
  dilatation~$(1+\sqrt{2})^2$.}
\label{fig:braidpicture}
\end{figure}

Now we introduce a convenient way to describe rod motions: braids.
Picture a given two-dimensional rod motion as a three-dimensional
space-time plot, as in Fig.~\ref{fig:taffybraid}, where the vertical
axis represents time.
The resulting tangle of strands is called a \emph{physical braid}.
Two physical braids are equivalent if they can be deformed into each
other with no strands crossing other strands or boundaries.

\begin{figure}
  \begin{center}
  \subfigure[]{
    \includegraphics[height=.3\textheight]{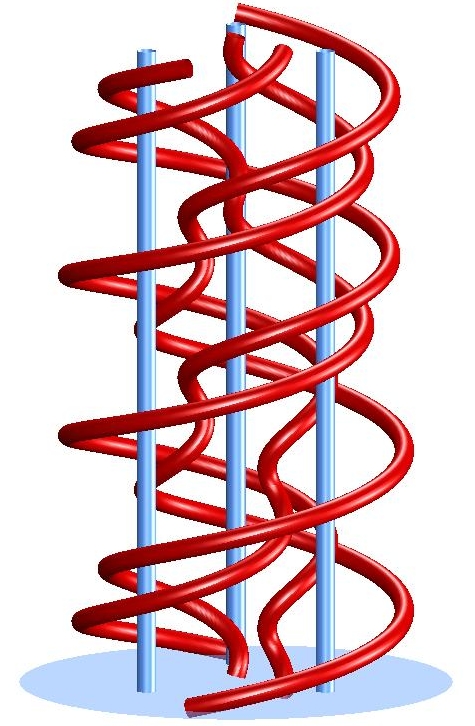}
    \label{fig:breadbraidfixed}
  }\hspace{.1\textwidth}%
  \subfigure[]{
      \includegraphics[height=.3\textheight]{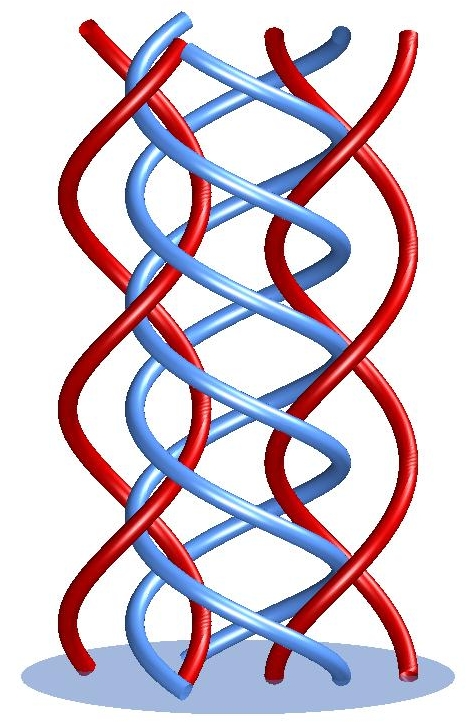}
      \label{fig:breadbraid}
  }
  \end{center}
  \caption{(a) The trajectories of the seven rods of the mixograph in
    Fig.~\ref{fig:pinmixer} viewed in a frame of reference where the
    three lower baffles are fixed. (b) The same motion, but viewed in
    a rotating frame where the four upper pins rotate in circles and
    the lower mixing vat and baffles counter-rotate in the opposite
    direction.  (Three periods are shown.)  In (b) the rod paths
    repeat the $\nn=7$ strand braid $\sigma_{3} \sigma_{2} \sigma_{3}
    \sigma_{5} \sigma_{6}^{-1} \sigma_{2} \sigma_{3} \sigma_{4}
    \sigma_{3} \sigma_{1}^{-1} \sigma_{2}^{-1} \sigma_{5}$. The
    dilatation associated with this braid is approximately~$4.1858$,
    and its topological entropy is~$1.4317$.}
\label{fig:breadbraids}
\end{figure}

An example of a device whose rod motion can be described by a physical
braid is shown in Fig.~\ref{fig:taffya}, which is a photograph of a
\emph{taffy puller}.  (The purpose of the taffy puller is to stretch
and fold candy to oxygenate it, making it lighter and chewier.)  The
central rod is fixed, and two rods orbit the central one following a
motion shown in Fig.~\ref{fig:taffyb} and as a space-time plot in
Fig.~\ref{fig:taffybraid}.  Fixed rods are often called
\emph{baffles}, though the distinction between fixed and moving rods
is immaterial to a topological description.  
We will analyse the taffy puller in more detail in the next section.

The physical braid describing the rod motion of the mixograph in
Fig.~\ref{fig:pinmixer} is a little more complicated. The motion may
be viewed in a frame of reference where either the three lower baffles
are fixed (Fig.~\ref{fig:breadbraidfixed}) or are rotating
(Fig.~\ref{fig:breadbraid}). The braids appear different in both
cases, but yield the same topological information about the complexity
of the flow.

We can pass from physical braids to an algebraic description of braids
by introducing \emph{generators}.  Assume without loss of generality
that all~$\nn$ stirrers (\ie, stirring rods) can be ordered from left
to right along some projection line.  The generator $\sigma_i$,
$i=1,\dotsc,\nn-1$, denotes the clockwise interchange of the $i$th rod
with the $(i+1)$th rod along a circular path, where~$i$ is the
position of a rod counting from left to right.  The anticlockwise
interchange is denoted~$\sigma_i^{-1}$ (see
Fig.~\ref{fig:taffybraid_diagram} for a depiction of generators.)  We
can write consecutive interchanges as a sequence, for
example~$\sigma_1\sigma_{2}^{-1}$, where the generators are read from
left to right (this convention differs from \citet{Boyland2000}).
This composition law allows us to define the \emph{Artin braid
  group on $\nn$ strands}, $\Br_\nn$, with the identity given by
untangled strands.  A sequence of generators is thus an element
of~$\Br_\nn$ and is called a \emph{word} or \emph{braid word}.
(Repeated generators are written as powers, such
as~$\sigma_1\sigma_1=\sigma_1^2$
and~$\sigma_2^{-1}\sigma_2^{-1}=\sigma_2^{-2}$.)  Braids are often
written in a standard form as a \emph{braid diagram}, as in
Fig.~\ref{fig:taffybraid_diagram}.

In order that braid
group elements correspond to physical braids, they must satisfy the
additional relations~\citep{Birman1975}
\begin{subequations}
\begin{alignat}{2}
\sigma_i \sigma_{j} \sigma_i &= \sigma_{j} \sigma_{i} \sigma_{j} \qquad
&\mbox{if} \qquad |i-j| &= 1, \label{eq:121} \\
\sigma_i \sigma_j &= \sigma_j \sigma_i \qquad &\mbox{if} \qquad |i-j| &\ge 2.
\label{eq:comm}
\end{alignat}%
\label{eq:presentation}%
\end{subequations}%
These relations are sufficient to characterise~$\Br_\nn$, a nontrivial
fact first proved by Artin~\cite{Artin1947}.

The crucial role of braids as an organising tool is that they
correspond to isotopy classes.  (There is a subtlety involving
possible rotations of the outer boundary, so that one usually speaks
of \emph{braid types} specifying isotopy classes rather than
braids~\citep{Boyland1994,Boyland2000}.)  We assign a braid to stirrer
motions by the diagrammatic approach employed in
Fig.~\ref{fig:braidpicture}, and to every braid we can assign stirrer
motions by reversing the process.  Exploring possible stirring
protocols is now reduced to the algebraic task of examining braids,
which is more suitable to optimisation.

\section{Computing the Topological Entropy of a Braid}
\label{sec:entropy}

As we saw in Section~\ref{sec:braids}, a periodic rod-stirring
protocol can be associated with a braid on~$\nn$ strands, where~$\nn$
is the number of rods.  Corresponding to this braid is a very useful
measure of mixing and complexity, the \emph{topological entropy} of
the braid.  The topological entropy of a braid is the minimum
stretching rate that the rod motion imparts on material lines in the
flow~\cite{Yomdin1987, Newhouse1988, Newhouse1993, Boyland2000,
  Thiffeault2006}.  This minimum stretching rate is due to material
lines being dragged along by the rods, since they are not permitted to
go through the rods (think of the children's game `cat's
cradle'~\cite{Thiffeault2009_catscradle}, where fingers play the role
of rods).  If a braid induces a mapping that is in a pseudo-Anosov
mapping class, then the topological entropy is positive.  This leads
to material lines growing exponentially, which is good for mixing as
it implies the interface between two solutes becomes more and more
convoluted with time.  The topological entropy is always non-negative,
and it forces exponential stretching on material lines only if it is
strictly positive.  An equivalent measure is the \emph{dilatation} of
a braid, which is simply the exponential of the entropy, and
conversely the entropy is the logarithm of the dilatation.

The topological entropy is only a lower bound on the rate of
stretching of material lines: the lines can develop convoluted
`secondary folds' which are not forced by rods.  This occurs, for
instance, if the material lines are not pulled tight around the rods,
but rather have loose sections that fold upon themselves.  In fluid
dynamics there is usually a gap between the lower bound and the
measured rate of stretching of material lines in the fluid.  In
applications such as bread dough mixing and taffy pulling the lower
bound is typically sharp, since the elastic nature of the material
tends to make it wind tightly around the rods.

The problem is then to compute the topological entropy and dilatation
of a given braid.  For three strands (\ie, three rods), the Burau
representation offers an efficient method of computing the
entropy~\citep{Burau1936,Fried1986,Kolev1989,BandBoyland2007}.  The Burau
representation associates with each generator a two-by-two matrix,
\note{These are the transpose of BAS, so we must read braids and
  multiply left-to-right rather than right-to-left.}
\begin{equation}
  [\sigma_1] = \begin{pmatrix}1 & 0\\ -1 & 1\end{pmatrix},\qquad
  [\sigma_2] = \begin{pmatrix}1 & 1\\ 0 & 1\end{pmatrix},
  \label{eq:Buraun=3}
\end{equation}
where the square brackets indicate the matrix representation of a
generator.  The representation for~$\sigma_1^{-1}$ and~$\sigma_2^{-1}$
is obtained by matrix inversion.  The Burau representation of a braid
word is then given simply by multiplying the corresponding matrices
together.  The spectral radius (the magnitude of the
largest eigenvalue) of the Burau representation of a braid word then
gives the dilatation of the braid~\citep{Burau1936, Fried1986,
  Kolev1989, BandBoyland2007}.  For example, for the braid
word~$\sigma_1^{-1}\,\sigma_2$, the Burau representation is
\begin{equation}
  [\sigma_1^{-1}\,\sigma_2] = [\sigma_1^{-1}]\cdot[\sigma_2] =
  \begin{pmatrix}1 & 0\\ 1 & 1\end{pmatrix}\cdot
  \begin{pmatrix}1 & 1\\ 0 & 1\end{pmatrix}
  = \begin{pmatrix}1 & 1\\ 1 & 2\end{pmatrix}.
\end{equation}
This matrix has spectral radius~$(3+\sqrt{5})/2$, which is the
dilatation, and hence the entropy is~$\log[(3+\sqrt{5})/2]$.  (As we
will discuss later, this dilatation is the square of the golden
ratio~$(1+\sqrt{5})/2$.)  Roughly speaking, the Burau representation
counts the number of windings of loops around the rods, so its
spectral radius measures the growth of these loops under repeated
application of the map.

As another example, we return to the taffy puller (Section \ref{sec:braids})
which executes the braid $\sigma_1^2 \sigma_2^{-2}$. The Burau representation
is
\[
[\sigma_1^2 \sigma_2^{-2}] = [\sigma_1]^2 \cdot [\sigma_2^{-1}]^{2} =
  \begin{pmatrix}1 & 0\\ -1 & 1\end{pmatrix}^2 \cdot
  \begin{pmatrix}1 & -1\\ 0 & 1\end{pmatrix}^2
  = \begin{pmatrix}1 & -2 \\ -2 & 5\end{pmatrix},
\]
which has the spectral radius $3 + 2 \sqrt{2} = (1+\sqrt{2})^2$, as 
presented above.

Computing the topological entropy using the Burau representation is
the simplest approach for three rods ($\nn=3$).  For more than three rods,
however, this approach only provides a lower bound.  Moreover, this
lower bound is often nowhere near sharp.  Hence, a reliable
computation of the entropy for~$\nn>3$ requires a different approach.
The most powerful algorithm is due to~\citet{Bestvina1995}: for a
given braid, the algorithm gives its isotopy class (\ie, finite-order,
pseudo-Anosov, or reducible) as well as its entropy.  The algorithm is
fairly complex, and a discussion of it is outside the scope of this
paper, but there is a C++ implementation written by Toby
Hall~\cite{HallTrain} that we use here.  When speed is important (as
when searching for optimal braids), we use the iterative algorithm of
\citet{Moussafir2006}, also described in \citet{Thiffeault2010}.

\section{Optimising over Generators}
\label{sec:optgen}

We have introduced in Section~\ref{sec:braids} the idea of
representing two-dimensional rod motions by a braid on~$\nn$ strands,
where~$\nn$ is the number of rods.  Then in Section~\ref{sec:entropy}
we discussed the dilatation and topological entropy of a braid, which
are lower bounds on the amount of stretching experienced by material
lines wrapped around the rods in each stirring period.  We now turn 
to the question of
optimisation: clearly some stirring protocols are better than others,
and we would like to know which ones are best.  We will take a
topological viewpoint, assuming that topological entropy is the
relevant quantity to optimise.  The other part is choosing a cost
function for our optimisation, and for this there are several choices.
We shall look at two in particular: topological entropy per generator
(TEPG), and topological entropy per operation (TEPO), and find the
latter to be better suited to practical applications.

\subsection{Topological Entropy per Generator (TEPG)}
\label{sec:TEPG}

Entropy can grow without bound as the length of a braid increases, so
a proper definition of an optimal entropy requires a `cost' associated
with the braid.  An obvious measure of efficiency is to divide the entropy by the
smallest number of generators required to write the braid word.  For
example, the braid $\sigma_1^{-1}\,\sigma_2$ has
entropy~$\log[(3+\sqrt{5})/2]$ and consists of two generators.  Its
topological entropy per generator (TEPG) is
thus~$\tfrac{1}{2}\log[(3+\sqrt{5})/2]=\log[(1+\sqrt{5})/2]$.  The
question of the maximal TEPG for a given braid group~$\Br_\nn$ is
well-posed.

In~$\Br_3$, \citet{DAlessandro1999} proved that the maximal TEPG is
given by repeating the word~$\sigma_1^{-1}\sigma_2$.%
\footnote{This is a similar result to \citet{Blondel2005}, who compute
  the `joint spectral radius' of a collection
  of matrices using different methods.}
Thus, the maximal TEPG is~$\log[(1+\sqrt{5})/2]$, the logarithm of the
\emph{golden ratio} $\phi_1=(1+\sqrt{5})/2$.  In general, the
\emph{metallic means}\footnote{These are sometimes called \emph{silver
    means}, but that name is easily confused with the silver ratio or
  second metallic mean, which we shall use extensively in this paper.}
are defined by~$\phi_m \ldef \tfrac{1}{2}(m+\sqrt{m^2+4})$ (see
Appendix~\ref{apx:metallicmeans}).  We now summarise the results for
higher~$\nn$ (see Appendix~\ref{apx:proof} for the sketch of a proof,
aimed at the specialised reader):
\begin{itemize}
\item For~\hbox{$\nn=4$}, the
  braid~$\sigma_1^{-1}\sigma_2\sigma_3^{-1}\sigma_2$ has maximal TEPG
  of~$\log\phi_1$, the same TEPG as for~$\nn=3$.
\item For~\hbox{$\nn>4$}, all pseudo-Anosov braids in~$\Br_\nn$ have
  TEPG strictly less than~$\log\phi_1$.
\end{itemize}
(We first presented these results in~\cite{Thiffeault2006}, which were
also independently formulated by \citet{Moussafir2006}.)  In other
words, the highest TEPG of~$\log\phi_1$ is only achieved for~$\nn=3$
or~$\nn=4$.  Intuitively, the optimal braids for~$\nn=3$ and~$\nn=4$
involve a tight combination of three or four strands to achieve their
high entropy per generator.  A braid with more strands requires extra
generators to make it irreducible (and hence pseudo-Anosov), thereby
reducing the TEPG. 

For more than a few rods, the rod motions with optimal TEPG are not
very relevant practically, but we note that in $\Br_\nn$ we can get
arbitrarily close to a TEPG of~$\log\phi_1$ by, for example, repeating
the braid~$\sigma_1^{-1}\sigma_2$ a large number of times, followed by
a few generators to make the braid irreducible (pseud-Anosov).  The
TEPG of such a braid will converge towards~$\log\phi_1$ as
more~$\sigma_1^{-1}\sigma_2$ motions are added.

\subsection{Topological Entropy per Operation  (TEPO)}
\label{sec:TEPO}

We saw in Section~\ref{sec:TEPG} that the maximum topological entropy
per generator (TEPG) is equal to (for~$\nn\le 4$) or is uniformly
bounded above (for~$\nn> 4$) by~$\log\phi_1$,
where~$\phi_1=(1+\sqrt{5})/2$ is the golden ratio.  But for
applications, optimising the TEPG is not so useful.  This is because
each generator is counted separately, the assumption being that rods
are moved sequentially according to the generators.  In practice,
however, it is better to move as many rods at the same time as
possible.  From the point of view of braid words, this means that we
should count commuting adjacent generators as a single operation.  For
example, the braid $\sigma_1^{-1}\sigma_3^{-1}\,\sigma_2\sigma_4$
%\begin{equation}
%  \sigma_1^{-1}\sigma_3^{-1}\,\sigma_2\sigma_4
%  \label{eq:-1-324}
%\end{equation}
contains four generators, but the corresponding rod motions can be
performed in two operations, since the two motions
$\sigma_1^{-1}\sigma_3^{-1}$ can be performed together, followed
by~$\sigma_2\sigma_4$.  In this viewpoint we count the `cost' of the
braid $\sigma_1^{-1}\sigma_3^{-1}\,\sigma_2\sigma_4$
%~\eqref{eq:-1-324} 
as being two, not four.  Thus we are led to examine the
topological entropy per operation, or TEPO, which 
is a more physically relevant quantity to optimise than the TEPG.

\begin{figure}
\begin{center}
\includegraphics[width=.65\textwidth]{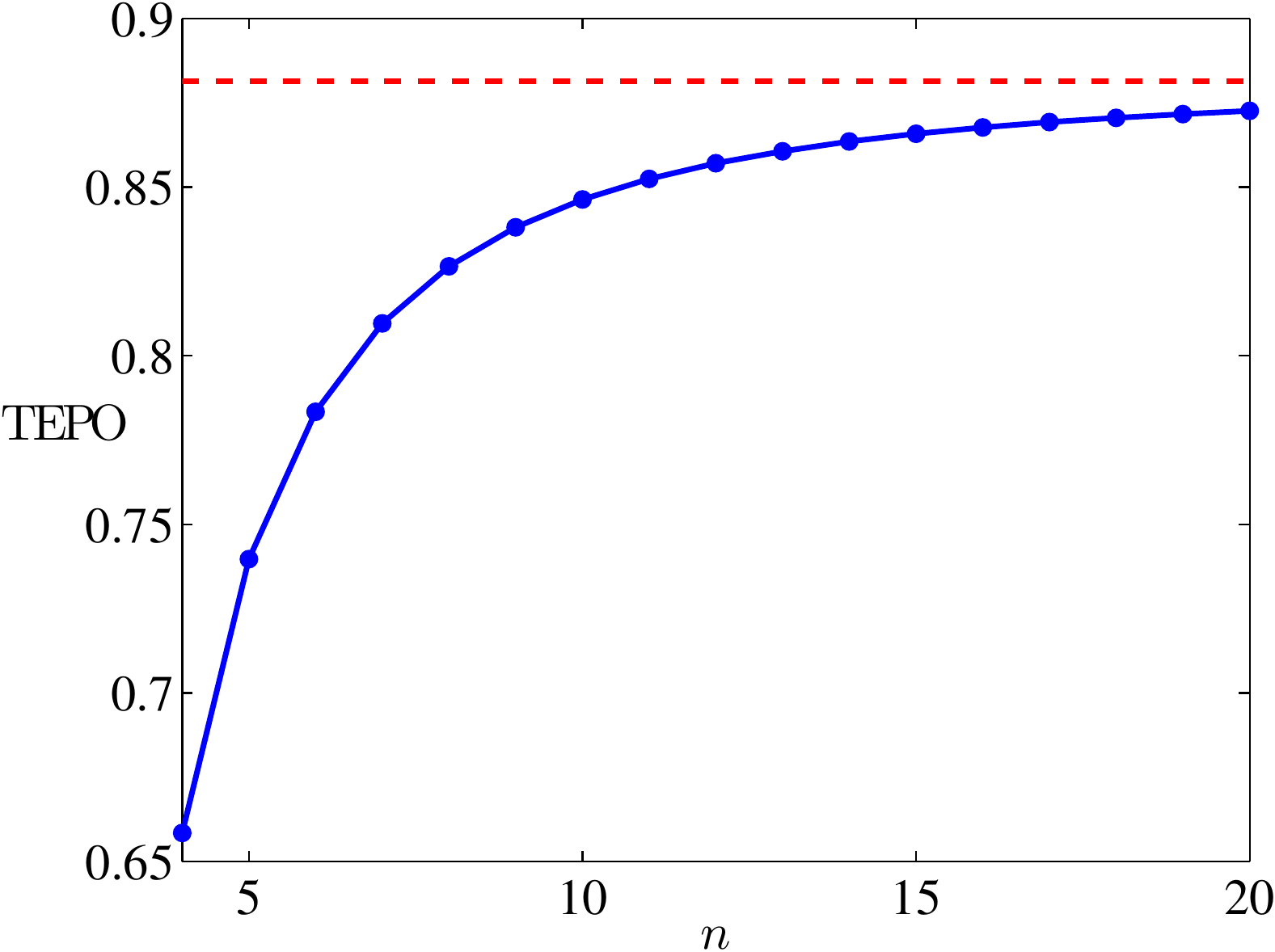}
\end{center}
\caption{Topological entropy per operation (TEPO) as a function
  of~$\nn$, the number of stirring rods.  The asymptote (dashed)
  is~$\log\phi_2\simeq0.8814$, where~\hbox{$\phi_2=1+\sqrt{2} \simeq
    2.4142$} is the silver ratio.}
\label{fig:opt_to_silver}
\end{figure}

Numerical investigation and the proof sketched in
Appendix~\ref{apx:proof} suggest that the optimal TEPO is given by the
braids
\begin{subequations}
\begin{gather}
  \sigma_1^{-1}\sigma_3^{-1}\cdots\sigma_{\nn-2}^{-1}\,
  \sigma_2\sigma_4\cdots\sigma_{\nn-1}, \qquad \text{for $\nn$ odd;}\\
  \sigma_1^{-1}\sigma_3^{-1}\cdots\sigma_{\nn-1}^{-1}\,
  \sigma_2\sigma_4\cdots\sigma_{\nn-2}, \qquad \text{for $\nn$ even.}
\end{gather}
\label{eq:maxTEPO}%
\end{subequations}
The braids given in~\eqref{eq:maxTEPO} consist of two operations:
interchange anticlockwise the first and second rod, third and fourth,
etc., then interchange clockwise the second and third rod, fourth and
fifth, etc.  These thus all have optimal TEPO because of the low
number of operations required.  In Fig.~\ref{fig:opt_to_silver} we
give the maximum TEPO as a function of~$\nn$: notice that it is
monotonically increasing, and it appears to asymptote to a fixed value
for large~$\nn$.  In Section~\ref{sec:annulus} and in
Appendix~\ref{apx:proof} we show that the asymptotic value
is~$\log\phi_2$, where~$\phi_2=1+\sqrt{2}$ is the \emph{silver ratio},
or the second metallic mean (see Appendix~\ref{apx:metallicmeans};
recall from Section~\ref{sec:TEPG} that the first metallic
mean~$\phi_1$ is the golden ratio).

\subsection{Other Types of Optimisation}

For practical applications, the type of optimisation we have discussed
has limited applicability.  The main issue is that the generators of
the braid group,~$\sigma_i$, do not necessarily correspond to natural
motions of rods in a given physical system.  For instance, the
exchange of the first and last rod in a three-rod system is
written~$\sigma_1\sigma_2\sigma_1$ in terms of exchanges of
neighbouring rods, so this simple operation requires three generators.
But clearly it does not cost much more energy to do this than to
exchange the first and second rods ($\sigma_1$).  The generators do
not capture the intrinsic geometry of the system.  For~$n=4$, the
exchange of the first and last rod requires five generators, making the
apparent cost even greater.

Many different optimisation problems can thus be formulated to
incorporate different engineering constraints.  In particular, the
resulting rod motions must be realisable using a straightforward
design.  We shall balance this engineering constraint with
mathematical tractability in Section~\ref{sec:silvermix}, where we
discuss a class of mixing devices based on braids on an annulus.

\section{Braids on an Annulus}
\label{sec:annulus}

In Section~\ref{sec:TEPO}, we selected the topological entropy per
operation (TEPO) as a suitable measure to optimise.  We found that the
optimal braid family given by~\eqref{eq:maxTEPO} had TEPO that
asymptotes to a fixed value for large~$\nn$
(Fig.~\ref{fig:opt_to_silver}).  In the present section we will find
the value of the asymptote by showing that the optimal braid for
large~$\nn$ can be realised as a braid on an annulus.

\begin{figure}
\begin{center}
\subfigure[]{
\includegraphics[width=.3\textwidth]{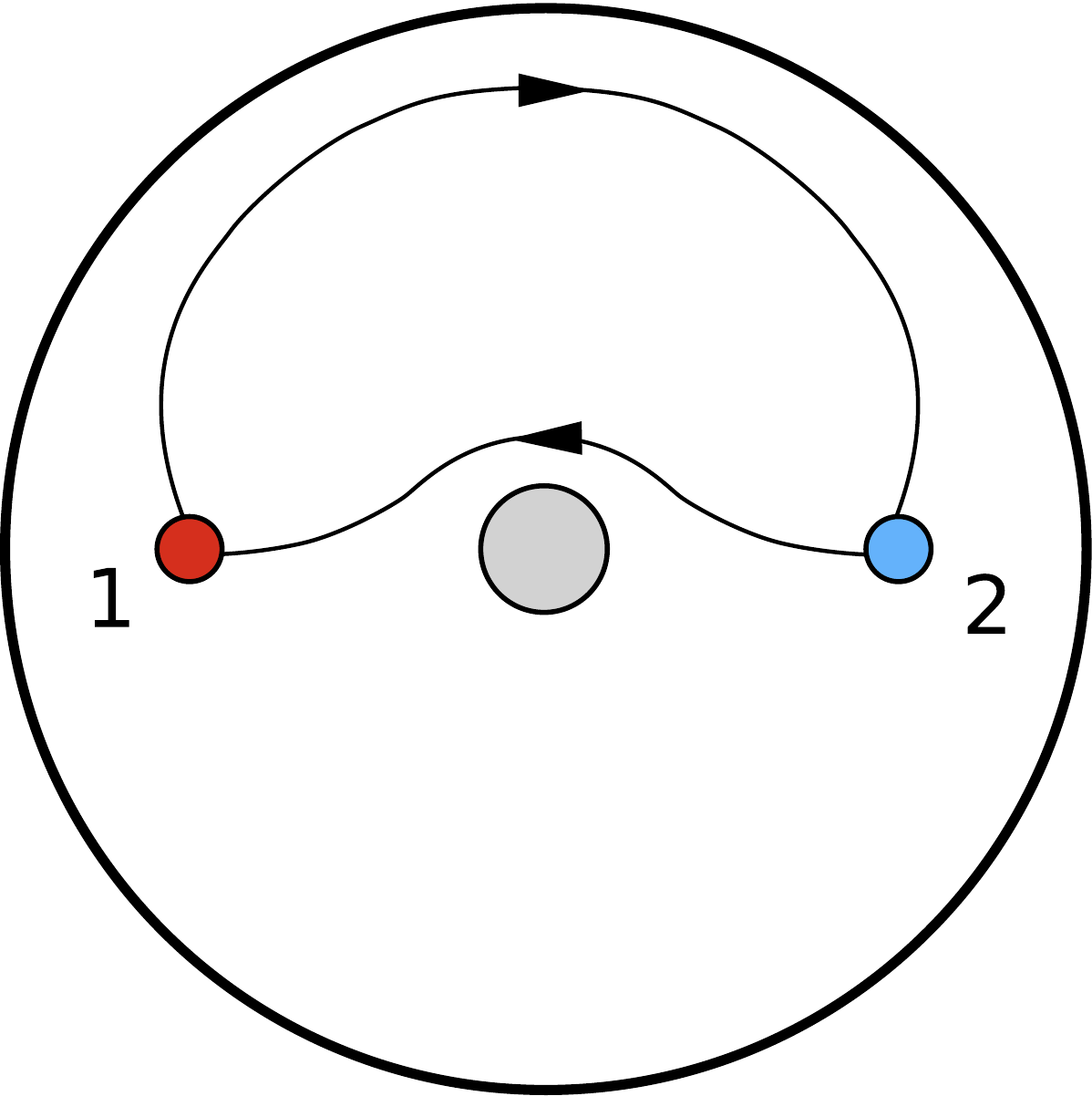}
\label{fig:Sigma1}
}\hspace{2em}
\subfigure[]{
\includegraphics[width=.3\textwidth]{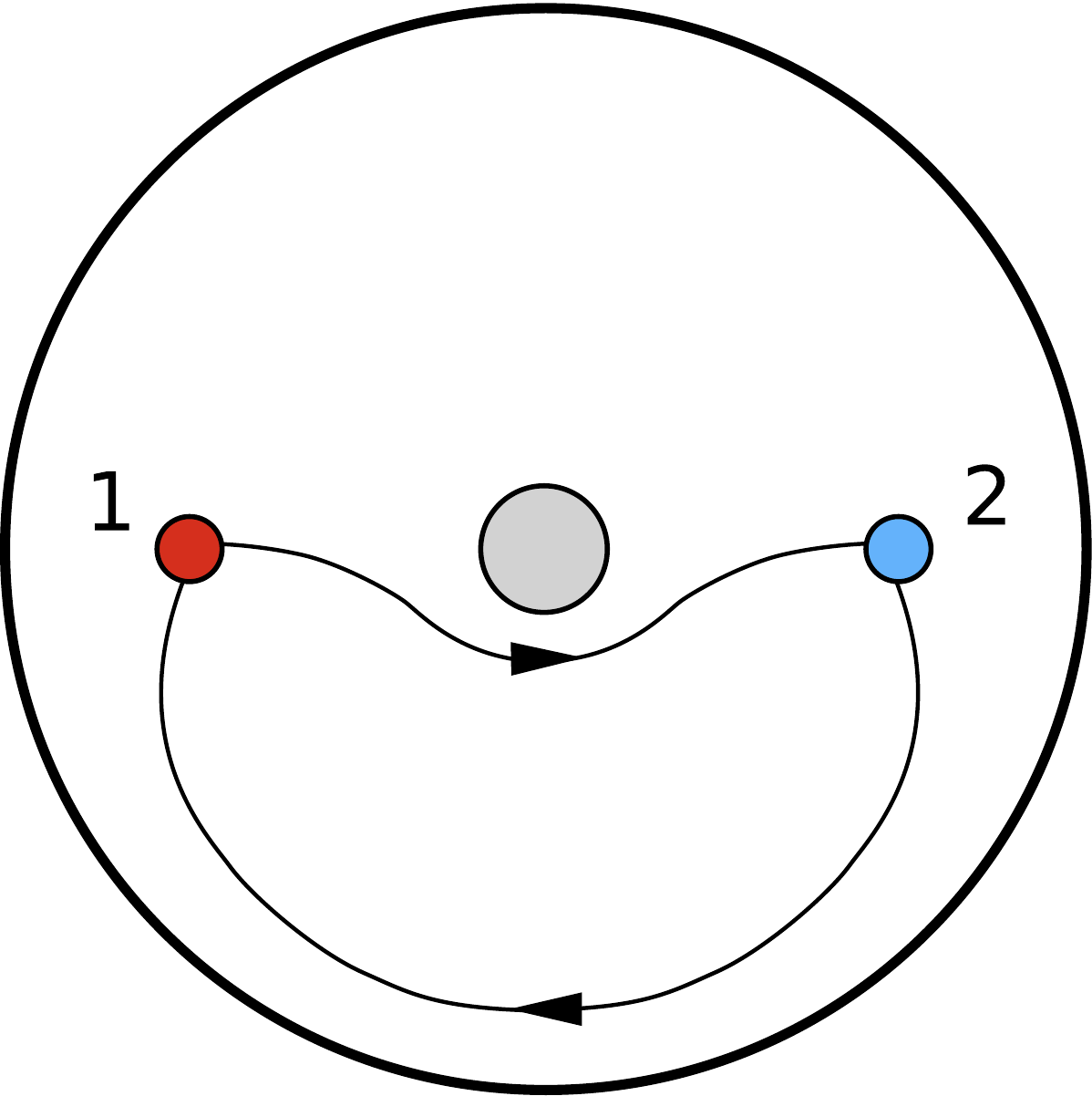}
\label{fig:Sigma2}
}\end{center}
\caption{Braid group generators for two rods in an annulus: (a)
  $\asigma_1$; (b) $\asigma_2$.  The central baffle remains fixed.}
\label{fig:Sigma}
\end{figure}

Consider two moving rods in an annular geometry, as shown in
Fig.~\ref{fig:Sigma}.
The central baffle 
%(larger in Fig.~\ref{fig:Sigma}) 
is fixed and
defines the annulus.  We define the operation~$\asigma_1$ as the
clockwise interchange of the two rods \emph{above} the central baffle
(Fig.~\ref{fig:Sigma1}), and~$\asigma_2$ as the clockwise interchange
of the two rods \emph{below} the central baffle
(Fig.~\ref{fig:Sigma2}).  These are the natural operations on an
annulus with two rods.  The generalisation to more rods is
straightforward.

To compute the entropy of braid words in this annular domain with two
rods and a central baffle, observe that we can also regard
Fig.~\ref{fig:Sigma} as a three-rod system, in which case we can
rewrite our two annular generators in terms of standard braid group
generators as
\begin{equation}
  \asigma_1 = \sigma_1\,\sigma_2\,\sigma_1^{-1}\,,\qquad
  \asigma_2 = \sigma_2\,\sigma_1\,\sigma_2^{-1}\,.
  \label{eq:asigmasigma}
\end{equation}
The standard braid group~$\Br_3$ has the well-known Burau
representation~\cite{Burau1936}, given by Eq.~\eqref{eq:Buraun=3}.  In
Section~\ref{sec:entropy} we explained that for~$\nn=3$ the spectral
radius of the Burau representation of a braid word gives the
dilatation of the braid.  We now use the
representation~\eqref{eq:Buraun=3} to derive a matrix representation
for~\eqref{eq:asigmasigma},
\begin{equation}
  [\asigma_1] = [\sigma_1]\,[\sigma_2]\,[\sigma_1^{-1}]
  = \begin{pmatrix}2 & 1\\ -1 & 0\end{pmatrix},\qquad
  [\asigma_2] = [\sigma_2]\,[\sigma_1]\,[\sigma_2^{-1}]
  = \begin{pmatrix}0 & 1\\ -1 & 2\end{pmatrix}.
  \label{eq:asigman=2}
\end{equation}
The spectral radius of the representation of an annular braid word
using~\eqref{eq:asigman=2} will thus give the exact dilatation of the braid.
To make the connection between the annular braid group and the Artin
braid group more explicit, we apply to~\eqref{eq:asigman=2} the
coordinate transformation \hbox{$[\asigma_i'] =
  R\,[\asigma_i]\,R^{-1}$}, with
\begin{equation}
  R = \begin{pmatrix}1 & 1 \\ -1 & 1\end{pmatrix}\,,
\end{equation}
yielding
\begin{equation}
  [\asigma_1'] = \begin{pmatrix}1 & 0\\ -2 & 1\end{pmatrix},\qquad
  [\asigma_2'] = \begin{pmatrix}1 & 2\\ 0 & 1\end{pmatrix}.
  \label{eq:asigman=2b}
\end{equation}
Note that although this is similar to the Burau representation of the Artin
braid group for~$\nn=3$, Eq.~\eqref{eq:Buraun=3}, it is not a valid
representation of that group since~$\asigma_1'\asigma_2'\asigma_1' \ne
\asigma_2'\asigma_1'\asigma_2'$.

Now if we define
\begin{equation}
  H = [\asigma_2'] = \begin{pmatrix}1 & 2\\ 0 & 1\end{pmatrix},
    \qquad
  V = [{\asigma_1'}^{-1}] = \begin{pmatrix}1 & 0\\ 2 & 1\end{pmatrix},
\end{equation}
we have recovered two matrices that are identical in form to~$H$
and~$V$ in \citet{DAlessandro1999}, with their parameter~$a=2$.  They
proved that the sequence that maximises the dilatation for a product
of matrices from the set~$\{H,V\}$, beginning with~$H$,
is~$HVHVHV\!\ldots$.  (For a sequence beginning with~$H$,
multiplication by~$H^{-1}$ or~$V^{-1}$ never increases the entropy as
much as multiplication by~$H$ or~$V$, basically because~$H^{-1}$
and~$V^{-1}$ involve negative matrix elements.)  Thus, the optimal
protocol with these generators consists of repeating
\begin{equation}
  HV = \begin{pmatrix}5 & 2\\ 2 & 1\end{pmatrix},
  \label{eq:HV}
\end{equation}
which has dilatation~$3+2\sqrt{2}=(1+\sqrt{2})^2=\phi_2^2$.  As
mentioned in Section~\ref{sec:TEPO}, the number~$\phi_2=1+\sqrt{2}$ is
the silver ratio.  For this reason, we call the
braid~$\asigma_2\asigma_1^{-1}$ corresponding to~$HV$ the \emph{silver
  braid}.

\begin{figure}
\begin{center}
\includegraphics[width=.75\textwidth]{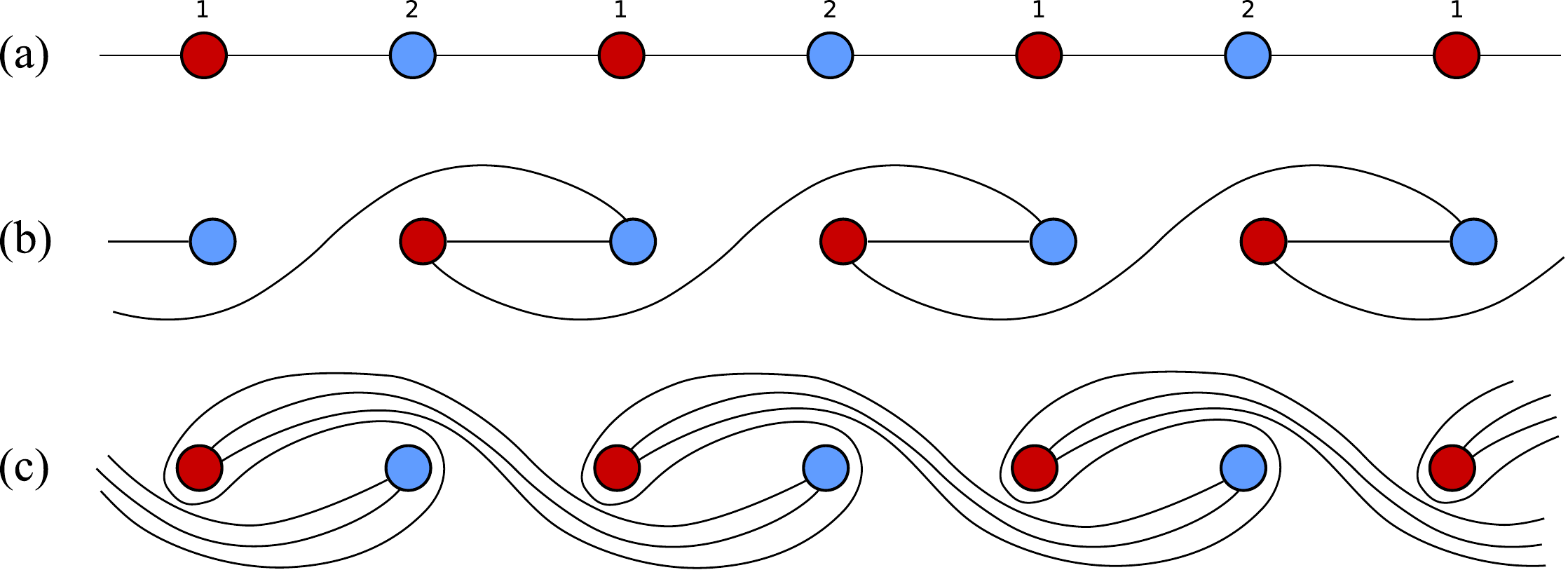}
\end{center}
\caption{(a) Two rods on a periodic lattice, joined by line segments. The
  segments are stretched by (b) the operation~$H=\Sigma_2$, a clockwise
  interchange of the rods at positions~$2$ and~$1$; and (c) $V=\Sigma_1^{-1}$,
  an anticlockwise interchange of the rods at position~$1$ and~$2$.
}
\label{fig:silver_lattice}
\end{figure}

Figure~\ref{fig:silver_lattice} shows a different way of looking at the
action of the braid~$\asigma_2\asigma_1^{-1}$.  Here we represent the
two rods of Fig.~\ref{fig:Sigma} on a periodic lattice, which is
topologically equivalent to the annulus.  (We can also regard the
periodic lattice as the universal cover of the annulus.)  In
Fig.~\ref{fig:silver_lattice}(b) we see the action of~$\Sigma_2$ on a
material line, followed by the action of~$\Sigma_1^{-1}$ in
Fig.~\ref{fig:silver_lattice}(c).  If we repeat the process, then the
length of the material line is stretched asymptotically by at least a
factor of~$\phi_2^2$ per application
of~$\asigma_2\asigma_1^{-1}$. Observe that we can implement a four-rod
protocol with the same dilatation by simply moving two pairs of rods
at once, a six-rod protocol by moving three pairs, etc.  It is this
scalability with number of rods that makes the annular braids
attractive: the optimal solution can be implemented with any even
number of rods.

Another great advantage of this scenario is that the silver
ratio~$\phi_2 \simeq 2.4142$ is considerably greater than the golden
ratio~$\phi_1 \simeq 1.6180$, so that material lines are stretched
much faster (at almost twice the rate).  A final desirable feature is
that an annular configuration is natural from an engineering
perspective, as we will see in Section~\ref{sec:silvermix}.

Note that the silver braid having greater entropy per generator than
the golden braid does not contradict the optimality conjecture of
Section~\ref{sec:optgen}, since that applied to a bounded domain,
whereas here we have a periodic array of rods.  We have also examined
topological mixing in periodic and biperiodic
geometries in~\citep{MattFinn2006}.

\section{Silver Mixers}
\label{sec:silvermix}

The great advantage of the configuration of Section~\ref{sec:annulus}
is that these optimal silver braids are readily implemented with
rotating machinery, despite the apparently complicated braiding
motion.  The easiest way to do this is by using moving planetary gears
orbiting a central fixed gear. A horizontal arm and stirring rod
attached beneath each planetary gear traces out a cycloid
pattern. Within this pattern fixed baffles can be placed, and if the
gear ratios and arm lengths are chosen appropriately the moving rods
will execute the over-and-under motion with the fixed baffles and
produce the silver braid (Section~\ref{sec:annulus}).  We
call~\emph{silver mixers} stirring devices whose entropy is a multiple
of~$\log\phi_2$.  The taffy-pulling device of Fig.~\ref{fig:taffy} is
an example of a silver mixer.

\begin{figure}
\begin{center}
\subfigure[]{
\includegraphics[width=.28\textwidth]{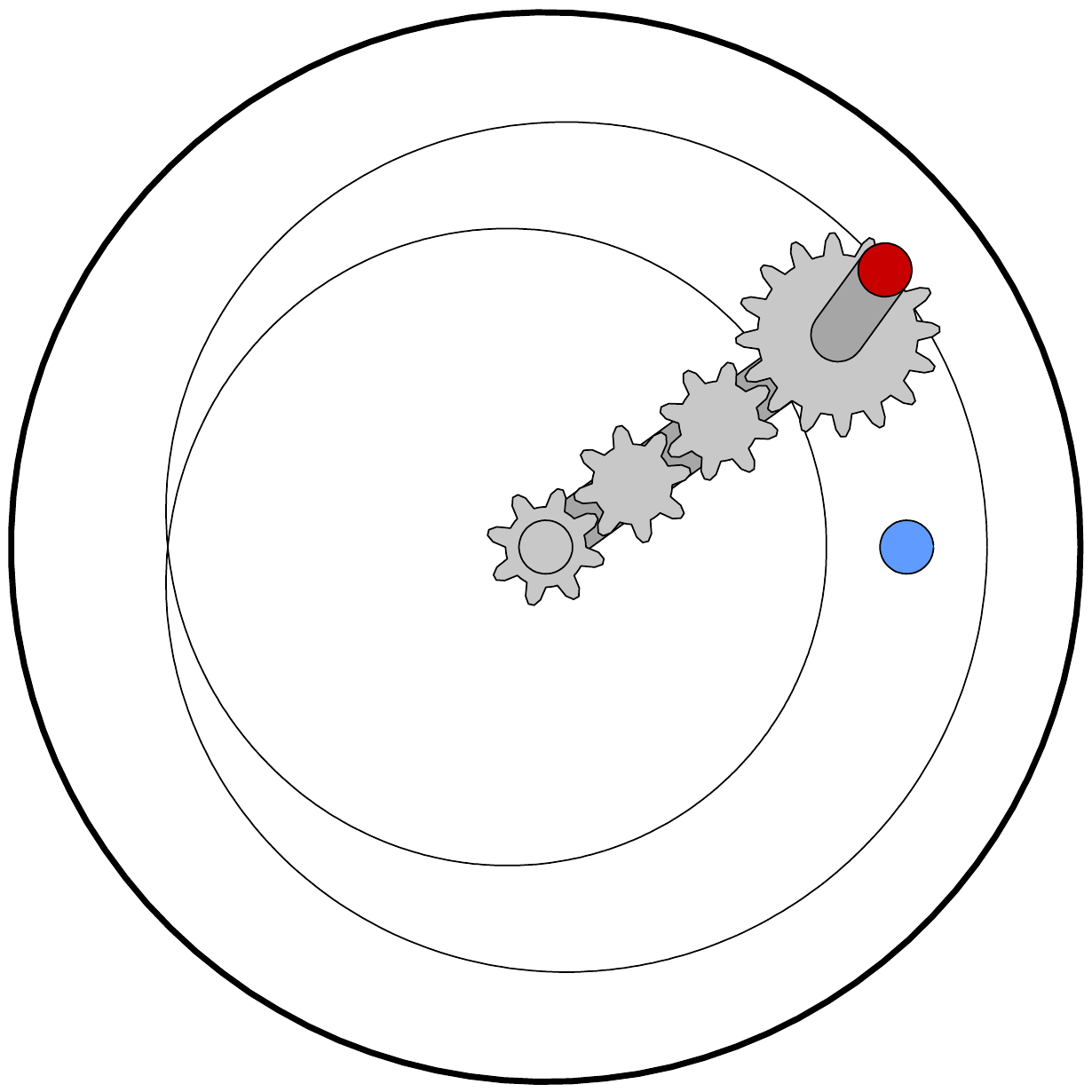}
\label{fig:silver02}
}\hspace{1em}
\subfigure[]{
\includegraphics[width=.28\textwidth]{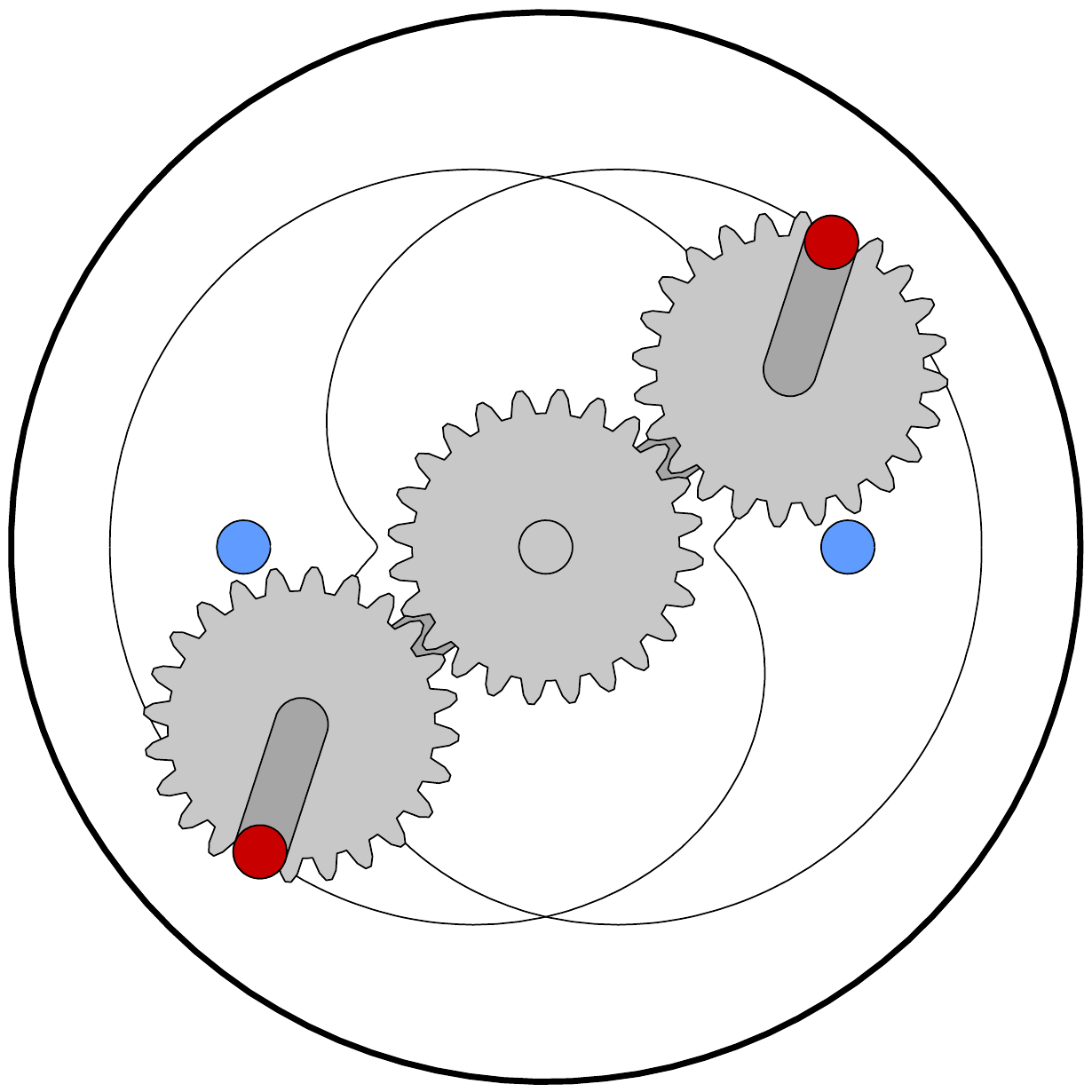}
\label{fig:silver04}
}\hspace{1em}
\subfigure[]{
\includegraphics[width=.28\textwidth]{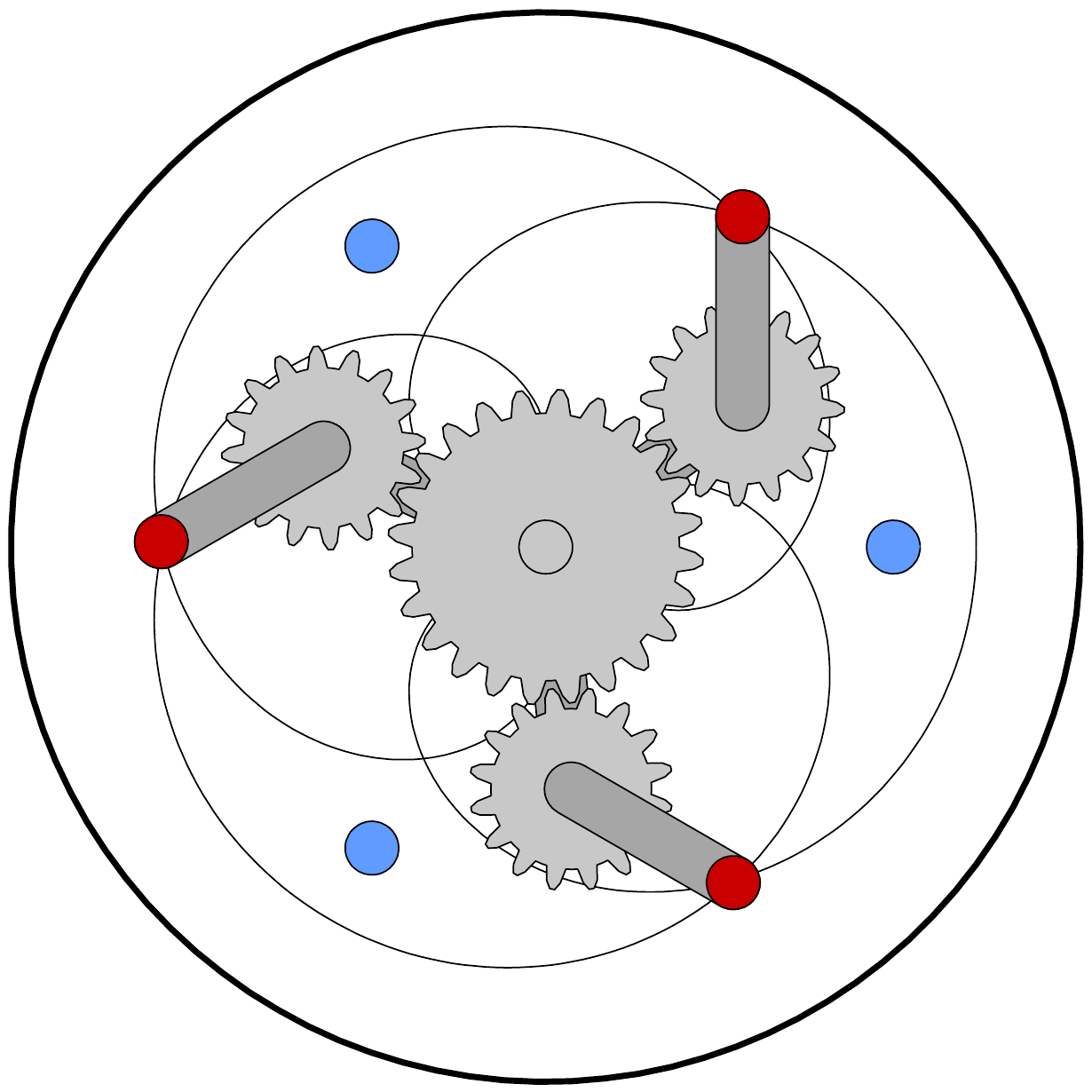}
\label{fig:silver06}
}\hspace{1em}%
\subfigure[]{
\includegraphics[width=.28\textwidth]{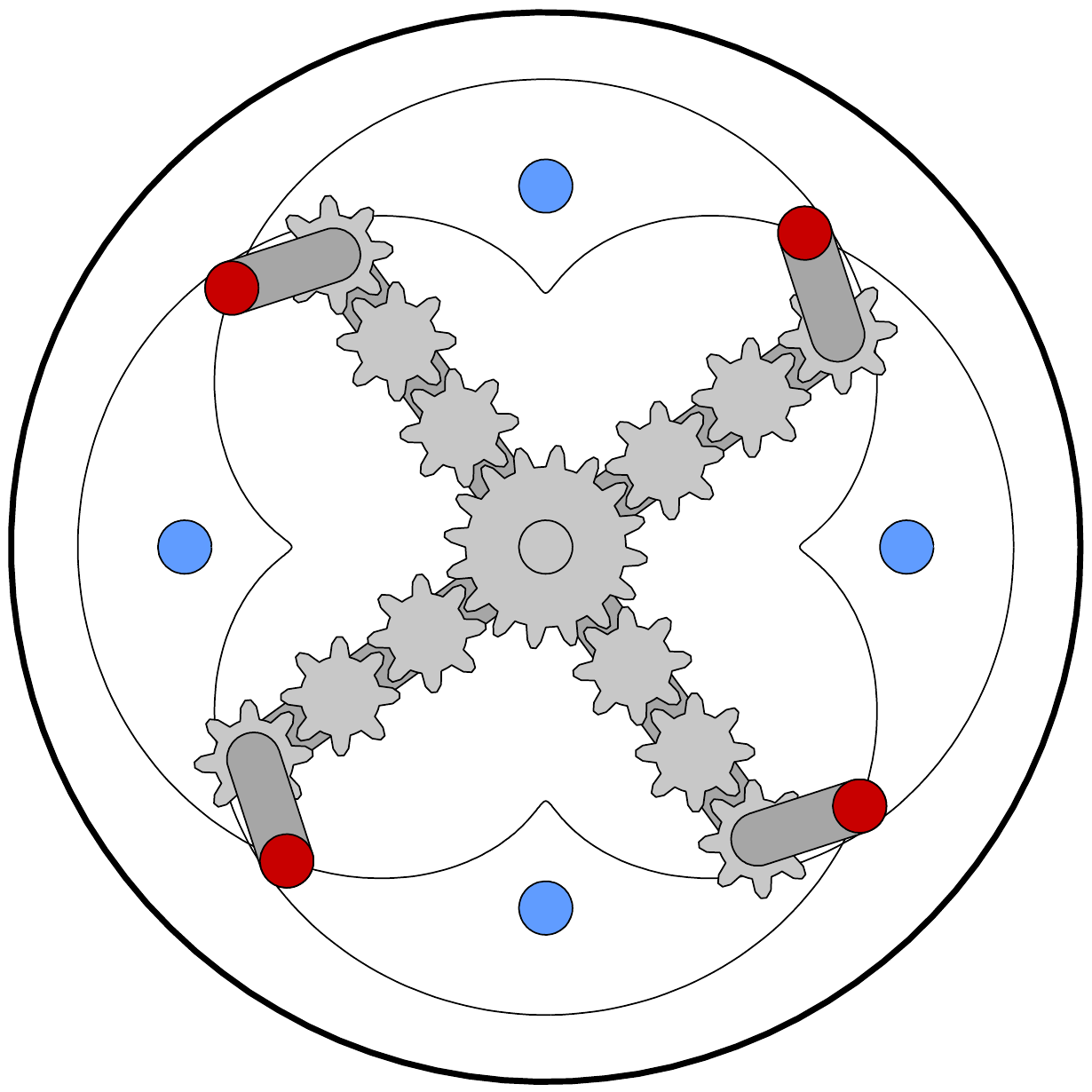}
\label{fig:silver08}
}\hspace{1em}
\subfigure[]{
\includegraphics[width=.28\textwidth]{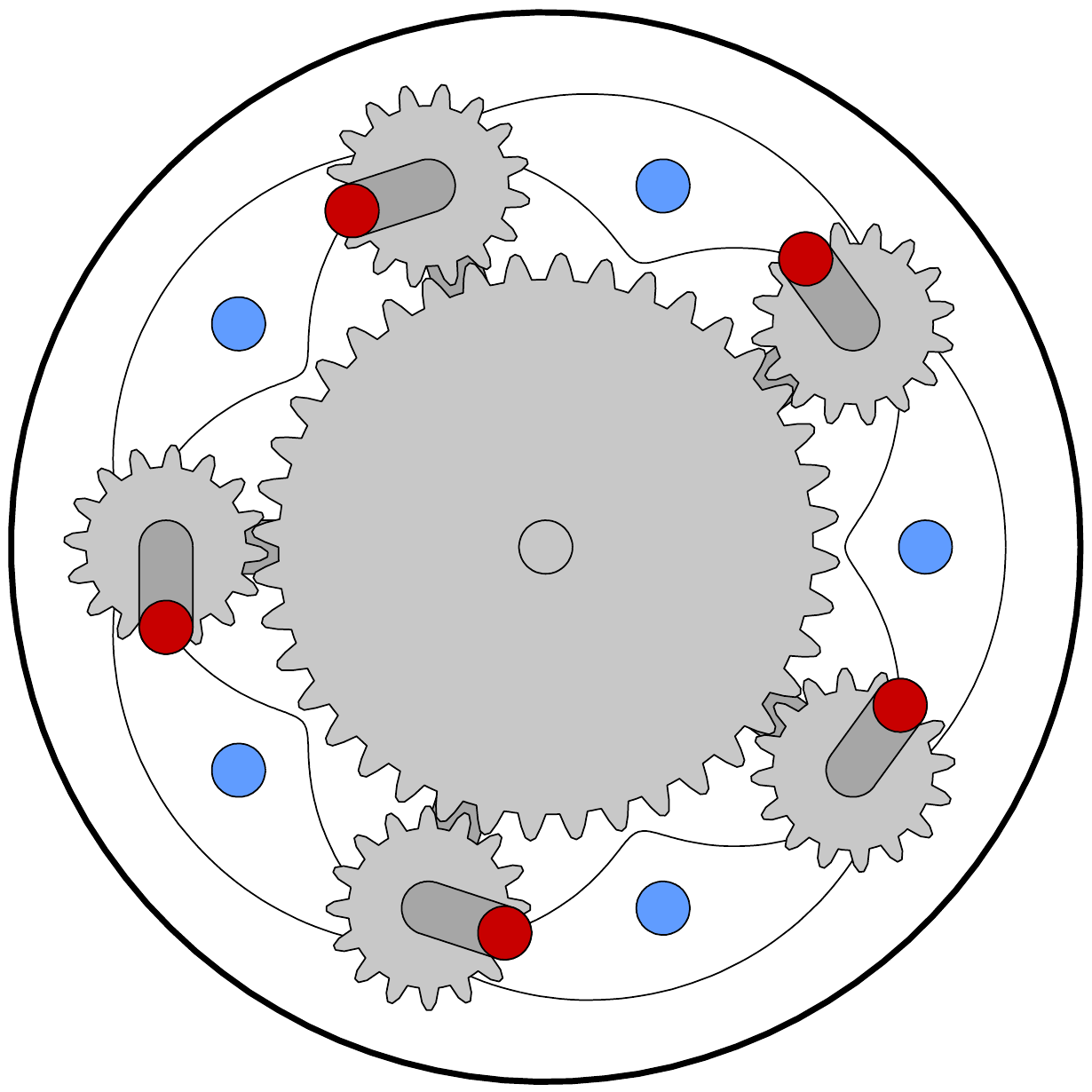}
\label{fig:silver10}
}\hspace{1em}
\subfigure[]{
\includegraphics[width=.28\textwidth]{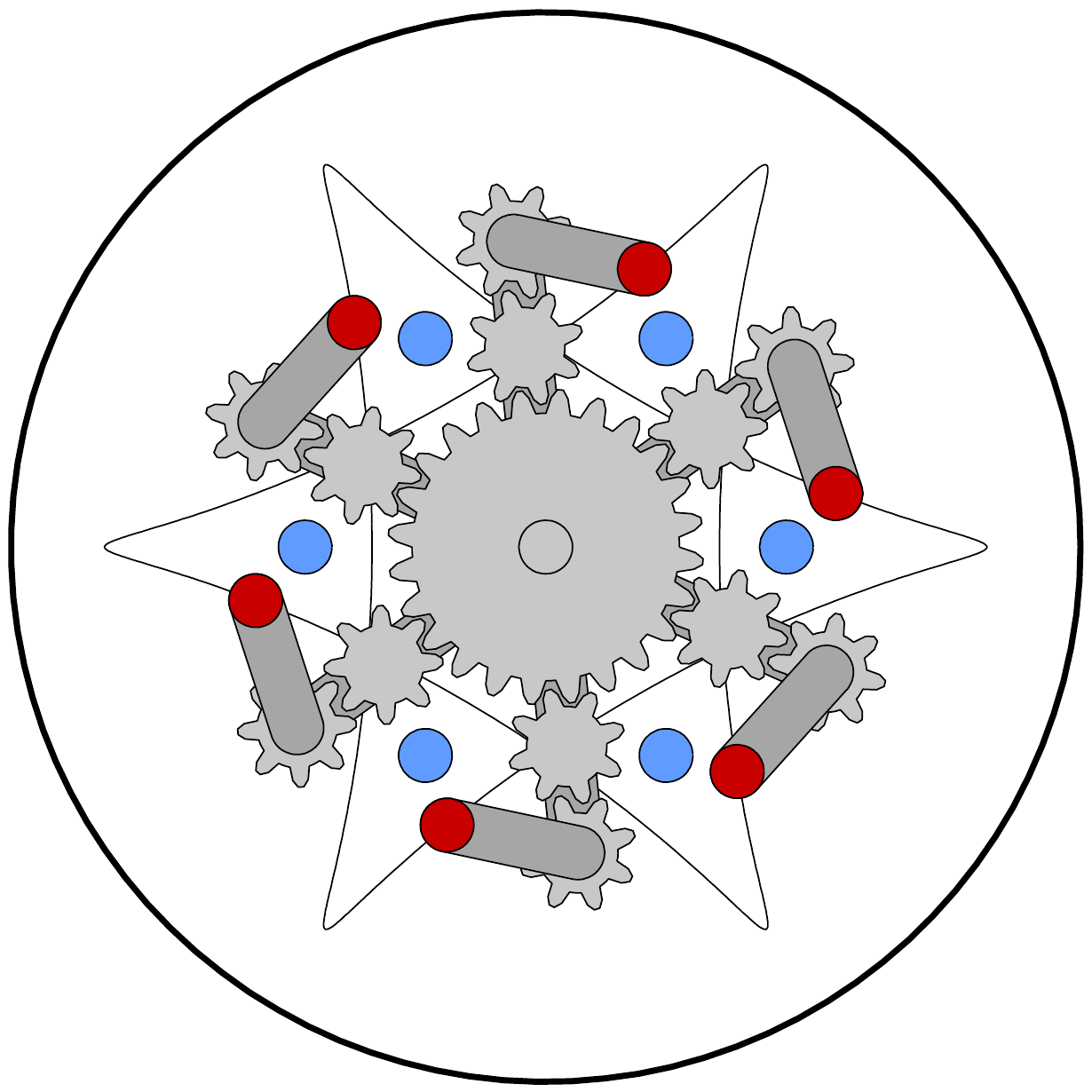}
\label{fig:silver12}
}
\end{center}
\caption{The first few silver mixers from two (top-left) to twelve
  (bottom-right) rods, excluding the central rod.  The paths of the
  moving rods are shown, as well as the arms connecting them to the
  planetary gears.  The device in~(a) is topologically identical to
  the taffy-puller of~Fig.~\ref{fig:taffy}.}
\label{fig:silver}
\end{figure}
\begin{figure}
\begin{center}
\subfigure[]{
\includegraphics[width=.28\textwidth]{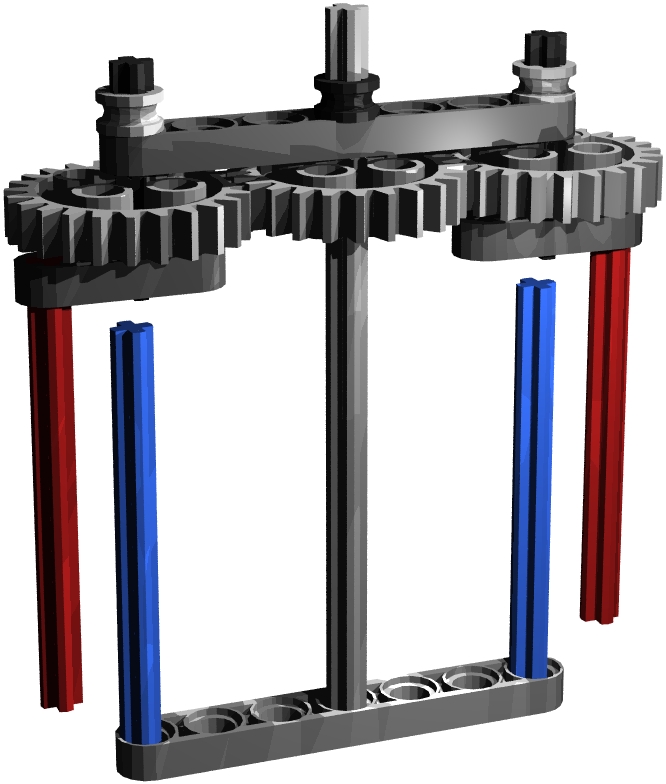}
\label{fig:lego_mixer5}
}\hspace{.1\textwidth}
\subfigure[]{
\includegraphics[width=.28\textwidth]{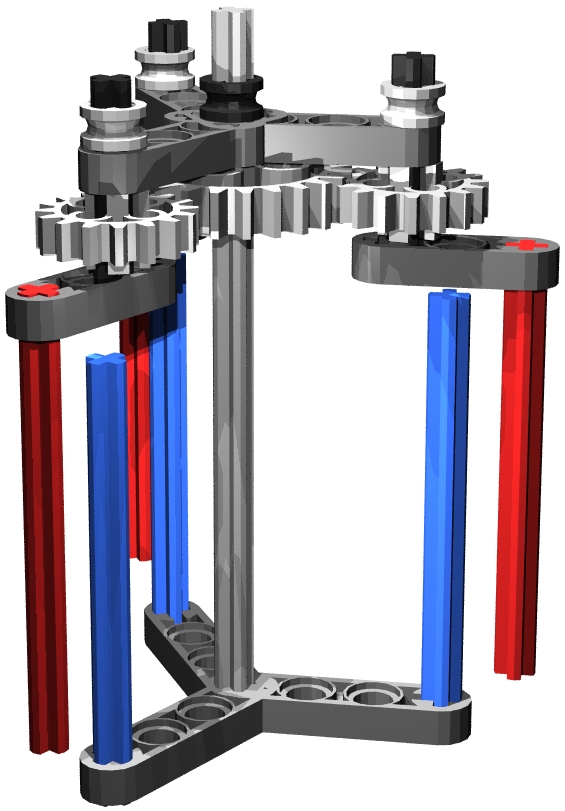}
\label{fig:lego_mixer7}
}
\end{center}
\caption{Implementation of the silver mixing protocols in
  Figs.~\ref{fig:silver04} and~\ref{fig:silver06} using
  Lego\texttrademark\ pieces~\cite{LegoCredits}.
}
\label{fig:lego_mixer}
\end{figure}
To illustrate the ease of construction, in Fig.~\ref{fig:silver} we
depict gear and arm arrangements which produce silver braids using
ordinary gearing available in toy Technic Lego\texttrademark. The red
(dark) rods are attached to the moving planetary gears, while the blue
(light) ones are fixed. A central rod is also employed, indicated in
grey. Often the central rod plays no significant topological role,
and can be removed.  A good reason for retaining the central rod,
however, is that it can support a lower set of arms that hold the
fixed rods, as in Fig.~\ref{fig:lego_mixer}.  This convenient feature
means that all of the mixing apparatus is in a single unit, and it is
not necessary to use a custom-made mixing vessel with baffles fixed to
the base (as for the mixograph in Fig.~\ref{fig:pinmixer}).
(\citet{Kobayashi2007} have also constructed a pseudo-Anosov stirring
device out of blocks, but it is not optimal.)

We show only the first six in the family of silver mixers, but in
principle one could produce the corresponding mixers with
$14,16,\ldots$ rods (plus the central rod).  Such devices would be
impractical, needing large numbers of very small gears and requiring
small arms for the moving rods to prevent them from colliding with
each other.

\begin{figure}
\begin{center}
\subfigure[]{
\includegraphics[width=.28\textwidth]{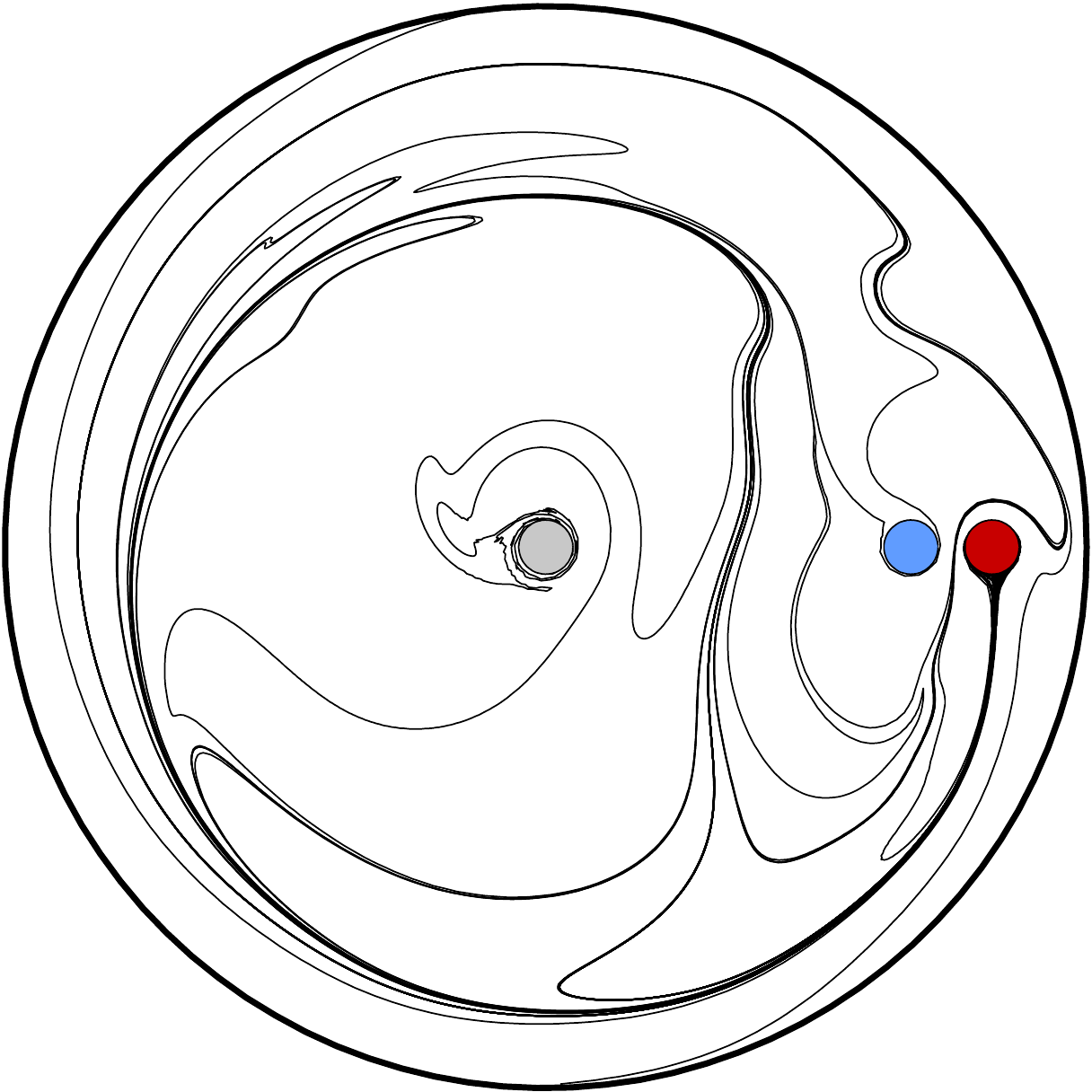}
\label{fig:silver02s}
}\hspace{1em}
\subfigure[]{
\includegraphics[width=.28\textwidth]{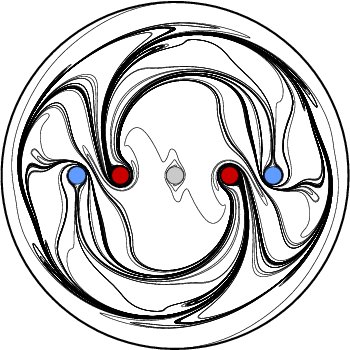}
\label{fig:silver04s}
}\hspace{1em}
\subfigure[]{
\includegraphics[width=.28\textwidth]{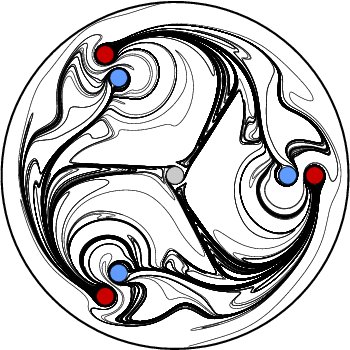}
\label{fig:silver06s}
}\hspace{1em}%
\subfigure[]{
\includegraphics[width=.28\textwidth]{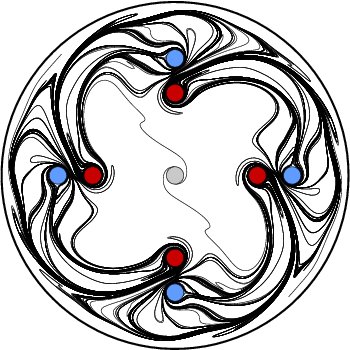}
\label{fig:silver08s}
}\hspace{1em}
\subfigure[]{
\includegraphics[width=.28\textwidth]{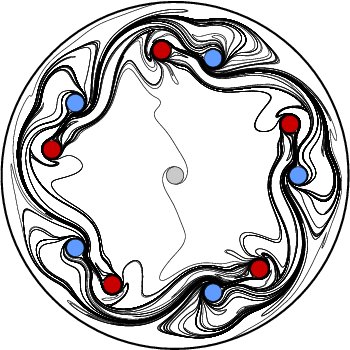}
\label{fig:silver10s}
}\hspace{1em}
\subfigure[]{
\includegraphics[width=.28\textwidth]{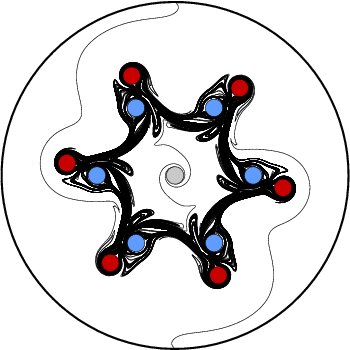}
\label{fig:silver12s}
}
\end{center}
\caption{Mixing patterns for the first few silver mixers, 
corresponding to the panels in Fig.~\ref{fig:silver}.}
\label{fig:silverpatterns}
\end{figure}

In Fig.~\ref{fig:silverpatterns} we present results from simulating
passive advection of a material line in the devices in
Fig.~\ref{fig:silver} in the Stokes flow regime, using the usual
particle insertion algorithms to maintain good resolution. (The
initial condition in each case being a vertical line from the top of
the mixer to the bottom.)  Each simulation ran until exponential
stretching overwhelmed computational resources, which in these
optimised devices happens very quickly.  For instance, in
Fig.~\ref{fig:silver12s} we were able to simulate only one complete
orbit of the planetary gears around the central rod.

If there are $\nn$ moving rods and $n$ fixed outer rods then one complete
orbit of a planetary gear around the fixed gear corresponds to 
$\nn/2$ motions $\asigma_2\asigma_1^{-1}$, and material lines are 
stretched by a factor of $(1+\sqrt{2})^{\nn/2}$. (A silver braid is 
completed every two orbits when $\nn$ is odd.) If the period of 
the flow is defined by a single orbit of the planetary gears then 
the entropy can be increased by including further pairs of rods, but only at
the expense of more complicated machinery and the additional energy input
associated with lots of rods moving in close proximity.

Of course, topological entropy is not the only consideration when
designing a mixing device. A large mixing region is also desirable,
but to determine the size of the mixing region one must solve the
particular dynamical equations governing the flow. The exact
dimensions of the gears, arms and mixing rods will also have a
significant effect on other mixing measures, and these could be tuned
by further simulation, provided that the apparatus still produces a
silver braid. We have made no attempt to optimise other mixing
measures here, since this introduces a wide range of other factors
depending on the specific application.  We note that in each
simulation in Fig.~\ref{fig:silverpatterns} the region of good
stretching is commensurate with the extent of the paths of the rods,
as is common in Stokes flow mixers~\cite{MattFinn2003}.

\begin{figure}
\begin{center}
\includegraphics[width=0.5\textwidth]{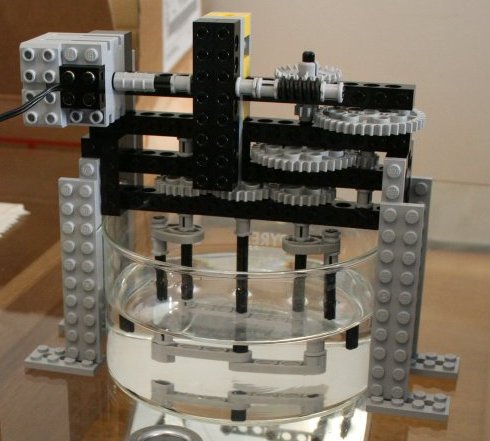}
\end{center}
\caption{Photograph of the experimental apparatus, corresponding to
  Figs.~\ref{fig:silver04} and~\ref{fig:lego_mixer5} in a rotating frame
  of reference in which all the rods are moving. Note that the extra
  gearing above the row of central and planetary gears is to
  reduce the speed of rotation to obtain slow (Stokes) flow.}
\label{fig:experiment}
\end{figure}

We close this section by presenting a comparison of simulations and
experimental results for a silver mixer similar to that in
Figs.~\ref{fig:silver04} and~\ref{fig:lego_mixer5}.  For convenience,
in this implementation we arranged for all four outer rods to rotate
(see Fig.~\ref{fig:experiment}).  The two planetary rods rotate in the
same direction in small circles around the centre of the planetary
gears, while the two `fixed' rods rotate around the central rod in the
opposite direction, supported from below by arms attached to the
central rod. In a rotating frame of reference this motion is identical
to that in Fig.~\ref{fig:silver04}, except for an additional rotation
of the outer boundary of the mixing vessel.

\begin{figure}
\begin{center}
\includegraphics[width=0.23\textwidth]{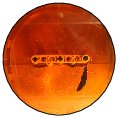}
\includegraphics[width=0.23\textwidth]{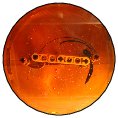}
\includegraphics[width=0.23\textwidth]{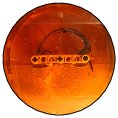}
\includegraphics[width=0.23\textwidth]{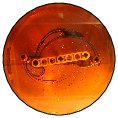}

\includegraphics[width=0.23\textwidth]{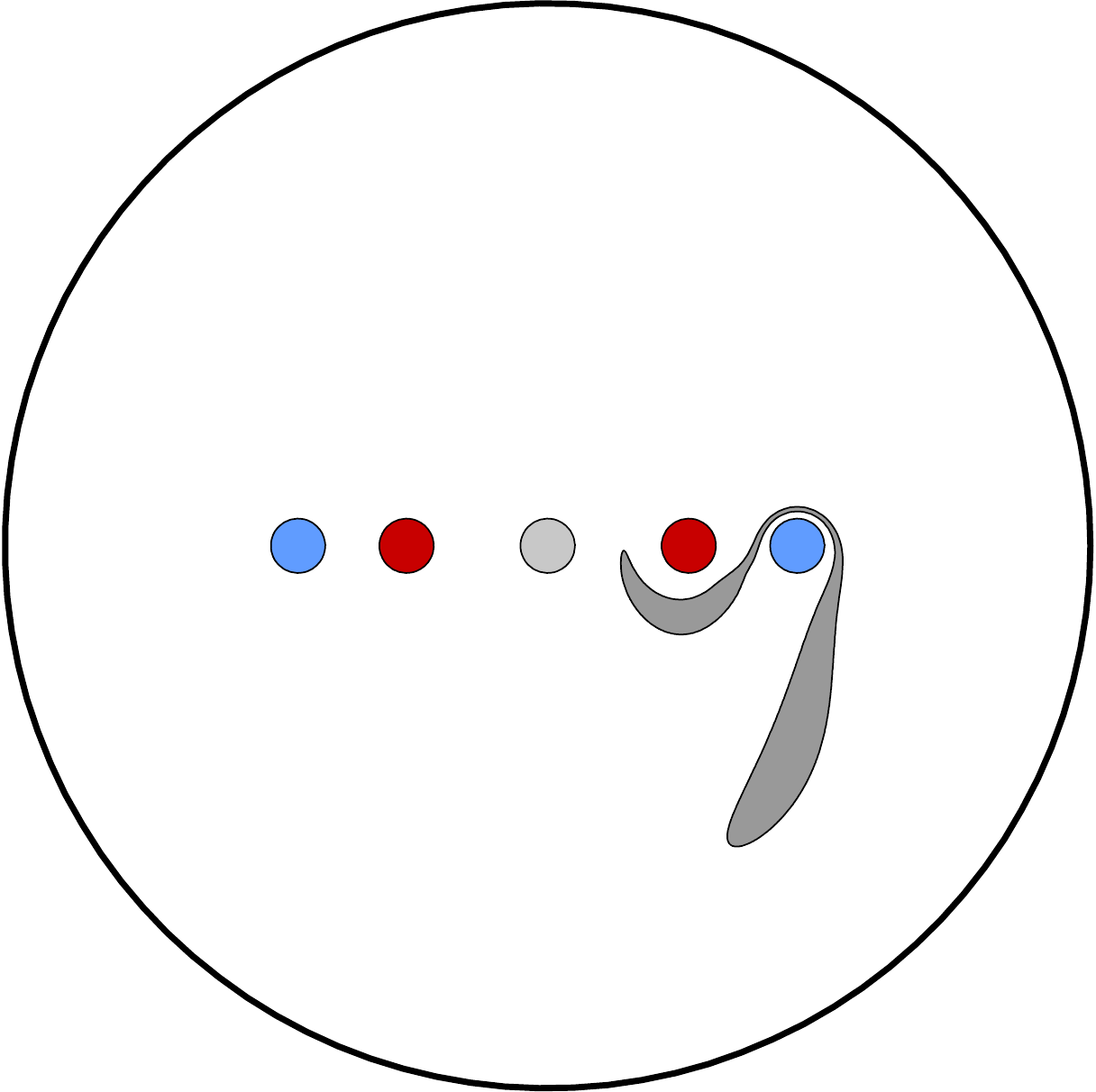}
\includegraphics[width=0.23\textwidth]{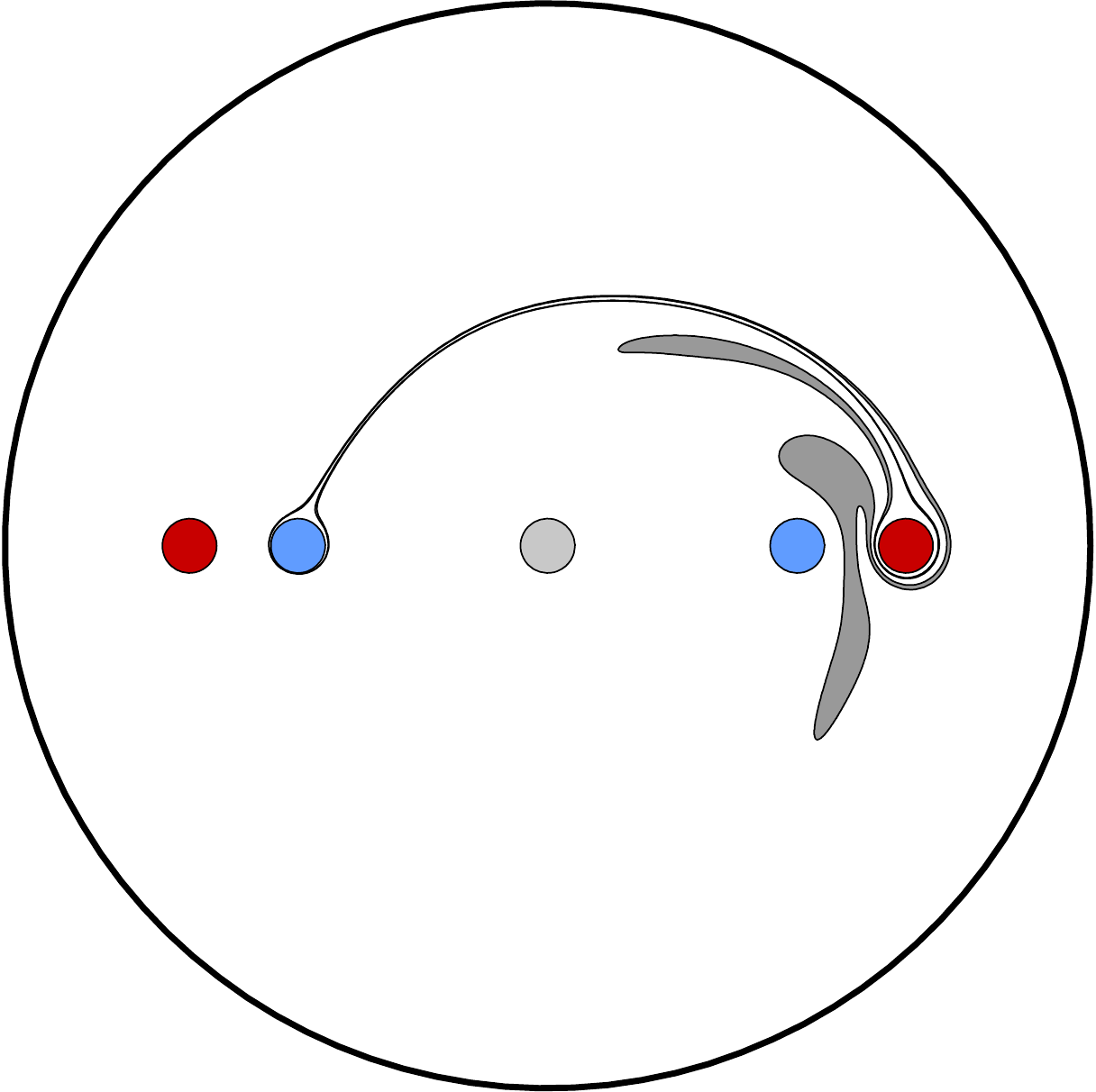}
\includegraphics[width=0.23\textwidth]{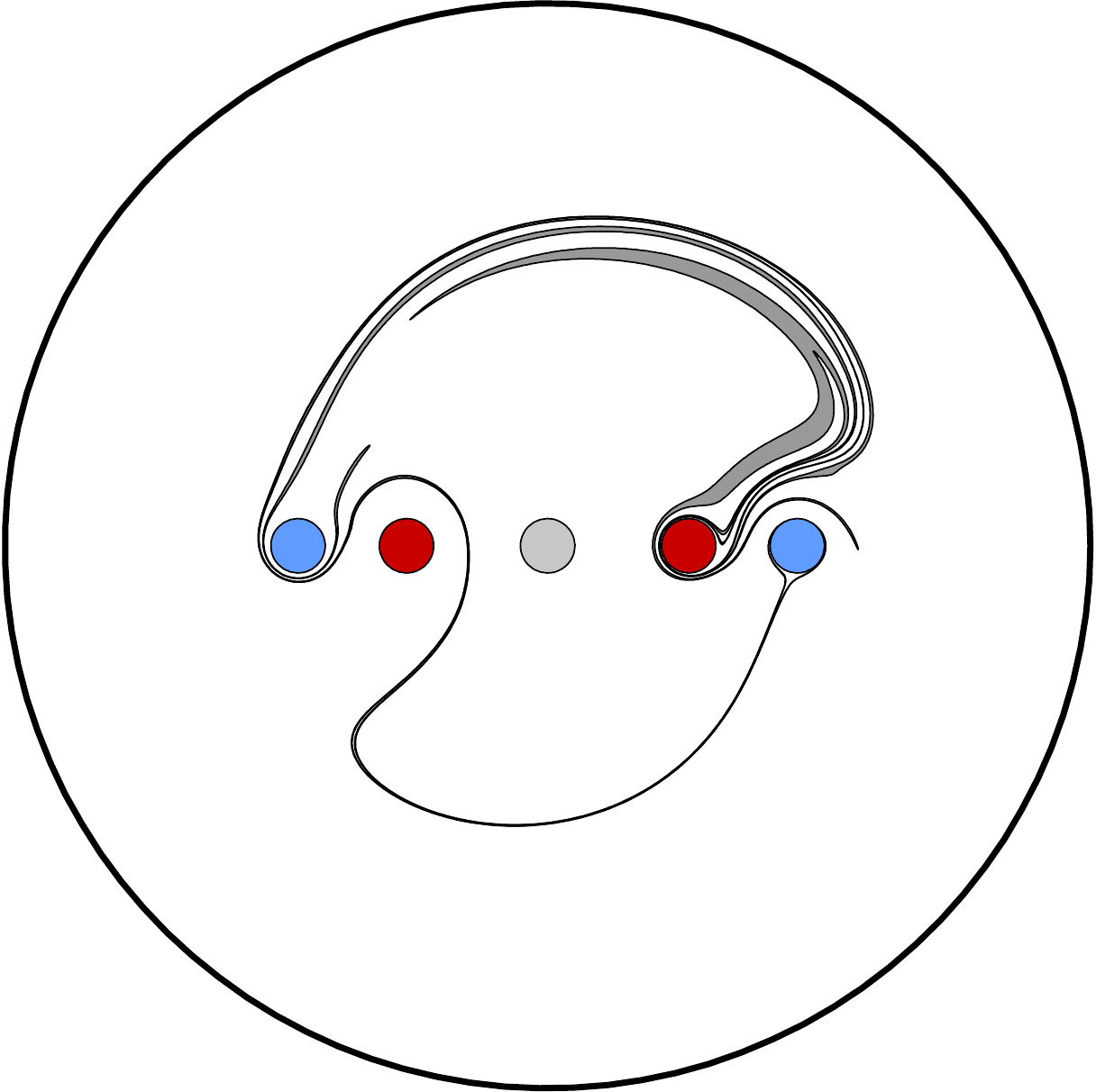}
\includegraphics[width=0.23\textwidth]{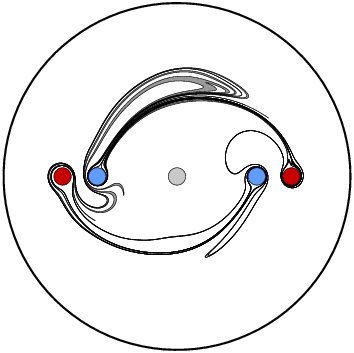}

\includegraphics[width=0.23\textwidth]{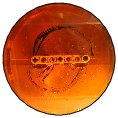}
\includegraphics[width=0.23\textwidth]{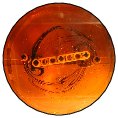}
\includegraphics[width=0.23\textwidth]{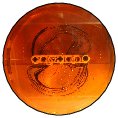}
\includegraphics[width=0.23\textwidth]{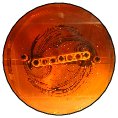}

\includegraphics[width=0.23\textwidth]{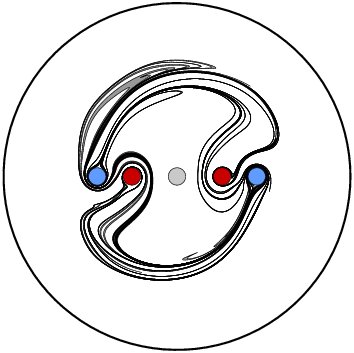}
\includegraphics[width=0.23\textwidth]{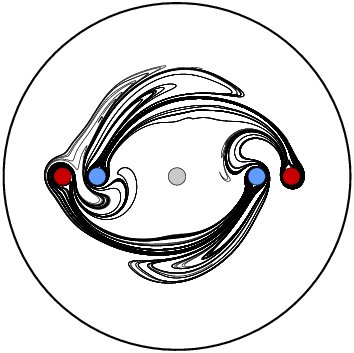}
\includegraphics[width=0.23\textwidth]{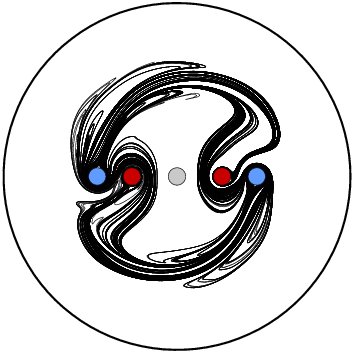}
\includegraphics[width=0.23\textwidth]{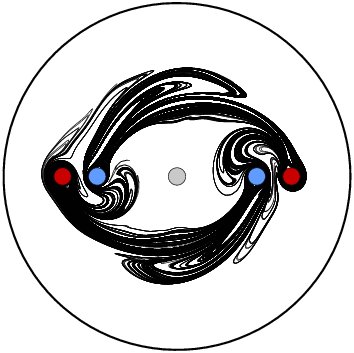}
\end{center}
\caption{Comparison of experimental and computational snapshots for
  the 4-rod silver mixer.  The patterns are virtually identical,
  indicating that three-dimensional effects are negligible, in spite of
  the supporting horizontal Lego rod spinning at the bottom of the
  mixing vessel.}
\label{fig:compare}
\end{figure}

The fluid used for the experiments was viscous golden syrup, and the
Stokes flow regime was attained by gearing down the motor so that a
single rotation of the rods would take several minutes (giving a
Reynolds number of order $10^{-4}$).  We used golden syrup mixed with
black food colouring as passive tracer, and tracked the evolution of a
circular blob over four complete rotations of the central rod. The
blob was photographed from below every half-rotation through the glass
worktop using a $45$-degree mirror, as shown in
Fig.~\ref{fig:compare}.

There is remarkably good agreement between the experimental pictures and
the corresponding numerical simulations, with surface and bottom
effects causing little discrepancy from the idealised two-dimensional
flow. It is readily observed that while the stretching is very rapid,
as it must be, the mixing pattern is qualitatively different from that
in Fig.~\ref{fig:silver04s}. This is due to the additional rotation of
the mixer boundary in a frame of reference rotating with the central
rod, which in this case appears to significantly reduce the extent of
the region in which the rapid stretching is observed.  Nonetheless,
the same high stretching rate is found in both mixers, as guaranteed
by the braiding motion.

\section{Discussion}
\label{sec:discussion}

In this paper, we discussed the optimisation of a common measure of
mixing quality---the topological entropy.  This is a crude measure
based on the motion of stirring rods, but it is relevant to many
fluid-dynamical situations, and crucial for pulling materials such as
bread dough or taffy.  The entropy is basically the rate of growth of
a hypothetical piece of dough wrapped around the rods.

Rod motions are best expressed in terms of the standard generators of
the braid group, which allows us to assign a `cost' to rod motion for
optimisation purposes.  We examined two such cost-normalised functions: the
topological entropy per generator (TEPG) and the topological entropy
per operation (TEPO).  The latter is less susceptible to assigning
high cost to relatively simple motions.  For a large number of rods,
the TEPG converges to the logarithm of the golden ratio, and the TEPO
converge to the logarithm of the silver ratio, though we do not have
rigorous proofs of these facts.

We then discussed optimisation using generators of the braid group on
the annulus for two rods, which is the natural limit of the TEPO for
an infinite number of rods.  This spatially-periodic configuration
lends itself well to mechanical construction.  We demonstrated this by
giving an explicit construction using simple gearing of stirring
devices with entropy given by multiples
of~$\log\phi_2=\log(1+\sqrt{2})$, where~$\phi_2$ is the silver ratio.
We also built such a device out of Lego\texttrademark\ pieces, as a
proof-of-concept.

The implementation of the silver mixers with planetary gearing
closely resembles the mixograph discussed earlier in Section \ref{sec:intro}
and shown in Fig.~\ref{fig:pinmixer}. However, the mixograph has four
moving rods but only three fixed baffles, so the rod motion lacks the 
rotational symmetry seen in the optimised silver braids. It is instructive
to compare the stretching produced by the mixograph with the silver mixer
shown in Fig.~\ref{fig:lego_mixer7} (since this device also has seven rods, 
including the central one). 

To make a fair comparison we note that the braid describing the
mixograph in Fig.~\ref{fig:breadbraid} is based on the shortest time
required for the rods to return to the same configuration in the
rotating frame of reference. This braid has an entropy of
approximately~$1.4317$.  Now consider the silver mixer in a rotating
frame where the rods on planetary gears rotate in circles and the
`fixed' rods counter-rotate. When the rods return to their initial
configuration exactly one silver braid has been completed, with
entropy~$1.7628$.  This is about~$23\%$ greater than the mixograph. In
this sense the mixograph is suboptimal, though the broken symmetry of
the rod arrangement improves other aspects of mixing, such as
uniformity.  For future work it would be desirable to define more
`universal' cost functions that are either less dependent on the
specific geometry, or instead better suited to particular engineering
applications.

\section*{Acknowledgements}

We thank Phil Boyland for helping us with the proof sketch in the
appendix, as well as Kiran Desai, Emmanuelle Gouillart, Toby Hall, and
Jacques-Olivier Moussafir for fruitful discussions.  J-LT was
partially supported by the Division of Mathematical Sciences of the US
National Science Foundation, under grant DMS-0806821.  MDF was
supported by the Australian Research Council, under Discovery Project
DP0881054.

\appendix

\section{Braids with Metallic Mean Dilatation}
\label{apx:metallicmeans}

From~\eqref{eq:Buraun=3}, the Burau representation for the braid
word~$\sigma_1^{-m}\sigma_2^m$ is
\begin{equation}
  \sigma_1^{-m}\sigma_2^m = \begin{pmatrix}1 & m \\ m & 1 + m^2\end{pmatrix}
  \label{eq:s1mms2m}
\end{equation}
with dilatation~$\phi_m^2$, where~$\phi_m$ are the so-called
\emph{metallic means}~\cite{silverratio}
\begin{equation}
  \phi_m \ldef \tfrac{1}{2}(m+\sqrt{m^2+4})\,.
\end{equation}
$\phi_1$ is the golden ratio, and~$\phi_2$ is known as the silver
ratio.  There is a simple geometrical construction of the metallic
means: start with a rectangle with one side of unit length, and
remove~$m$ unit squares.  The ratio of the sides of the remaining
rectangle is given by the~$m$th metallic mean if it is the same ratio
as the original rectangle.  The metallic means have continued fraction
representation~$\{m,m,m,m,\ldots\}$.

The three-rod protocol with dilatation~$\phi_m^2$ corresponds to~$m$
anticlockwise interchanges of rods~$1$ and~$2$, followed by~$m$
clockwise interchanges of rods~$2$ and~$3$.  Of course, not all braids
with metallic mean dilatation are of the
form~$\sigma_1^{-m}\sigma_2^m$; for instance, the annular
braid~$\asigma_1^{-1}\asigma_2$ (Eq.~\eqref{eq:HV}) has entropy equal
to~$\phi_2^2$.

\section{Optimal Topological Entropy: Sketch of Proof}
\label{apx:proof}

\renewcommand{\l}{\left}
\renewcommand{\r}{\right}
\mathnotation{\xpi}{x}
\mathnotation{\ypi}{y}
\mathnotation{\Kl}{K}
\mathnotation{\Klset}{\mathcal{K}}
\mathnotation{\Ll}{L}
\mathnotation{\Llset}{\mathcal{L}}
\mathnotation{\Lle}{\Ll_{\mathrm{even}}}
\mathnotation{\Llo}{\Ll_{\mathrm{odd}}}
\mathnotation{\cs}{c}
\mathnotation{\fib}{F}
\mathnotation{\as}{a}

Phil Boyland suggested that the methods used to obtain an upper bound
on the efficiency in his unpublished work on $\pi_1$-protocols would
also work for the topological entropy per generator (TEPG, see
Section~\ref{sec:TEPG}).  (The $\pi_1$ stirring protocols have a
single moving rod, with all the other rods fixed; they are called
$\pi_1$-protocols because their rod motions correspond to generators
of the fundamental group.)  We sketch a proof here, and extend it to
the TEPO.

Consider the action of a braid group generator~$\sigma_i$
($i=1,\dotsc,\nn-1$) on the standard generators~$\xpi_j$
($j=0,\dotsc,\nn$) of the fundamental group~$\pi_1$ of the disc
with~$\nn$ punctures.  (We include a generator~$\xpi_0$ around the
puncture corresponding to the disc's boundary when viewed as a
sphere.) This action is given by~\cite{Birman1975}
\begin{subequations}
\begin{alignat}{2}
  \xpi_i &\mapsto \xpi_i^{-1} \xpi_{i+1} \xpi_i\,, \nonumber \\
  \xpi_{i+1} &\mapsto \xpi_i\,, \nonumber \\
  \xpi_j &\mapsto \xpi_j\,, &\text{if } j &\ne i
  \text{ or } i+1. \nonumber \\
\intertext{%
Alternatively, we define the generators~$\ypi_j =
\xpi_j\cdots\xpi_0$, on which~$\sigma_i$ acts as
}
  \ypi_i &\mapsto \ypi_{i-1}\, \ypi_i^{-1}\, \ypi_{i+1}\,, \label{eq:action1}\\
  \ypi_j &\mapsto \ypi_j\,,\qquad\qquad\qquad &\text{if } j &\ne i.
  \label{eq:action2}
\end{alignat}
\end{subequations}%
The incidence matrix which just counts word length without
cancellations for~$\sigma_i$ is called $\Kl_i$.  This matrix has ones
on the diagonal and zeros elsewhere (from~\eqref{eq:action2}),
except~$(\Kl_i)_{i,i-1} = (\Kl_i)_{i,i+1} = 1$
(from~\eqref{eq:action1}).

For pseudo-Anosov mappings, the topological entropy is equal to the
growth rate of word length in~$\pi_1$~\cite{Bowen1978,Fathi1979} under
the action of the mapping.  This rate bounded above by the growth of
iterates of the corresponding products of incidence matrices.  The
maximal growth per generator is thus the log of the `joint spectral
radius' (JSR)~\cite{Rota1960} of the set of matrices~$\Klset=\{\Kl_i :
i=1,\dotsc,\nn-1\}$.  We
then have, for a given~$\nn$,
\begin{equation*}
  \text{TEPG} \le \log\JSR(\Klset)
  \le \log\max_i \lVert \Kl_i \rVert,
\end{equation*}
where~$\lVert\cdot\rVert$ is any induced norm.  Since the 1-norm of
all the~$\Kl_i$ is~$2$, it immediately follows that~$\text{TEPG} \le
\log 2 \simeq 0.6931$.  We can improve this by finding the exact JSR
of the matrices.  We first define the sup norm over all products
in~$\Klset$ of length~$k$,
\begin{equation*}
  \hat\rho_k(\Klset,\lVert\cdot\rVert)
  = \sup\l\{\l\lVert \prod_{m=1}^k M^{(m)}\r\rVert : M^{(m)} \in \Klset \r\}
\end{equation*}
and use the relation
\begin{equation*}
  \JSR(\Klset) =
  \limsup_{k\rightarrow\infty}\l(\hat\rho_k(\Klset,\lVert\cdot\rVert)\r)^{1/k}
\end{equation*}
which holds for bounded sets of matrices~\cite{Berger1992}.  We will
use the 1-norm~$\lVert\cdot\rVert_1$, which is the sup over
column-sums.  The form of the matrices~$\Kl_i$ means that a
matrix~$M^{(k)}$ with column-sums~$\cs_j^{(k)}$, $j=0,\dotsc,\nn$,
will after right-multiplication by~$\Kl_i$ have new column-sums
\begin{align*}
  \cs_{i-1}^{(k+1)} &= \cs_{i-1}^{(k)} + \cs_i^{(k)}\,, \\
  \cs_{i+1}^{(k+1)} &= \cs_{i}^{(k)} + \cs_{i+1}^{(k)}\,, \\
  \cs_j^{(k+1)} &= \cs_j^{(k)}\,, \qquad\qquad j \ne i-1 \text{ or } i+1.
\end{align*}
If we then right-multiply by~$\Kl_{i+1}$, we have
\begin{align*}
  \cs_{i}^{(k+2)} &= \cs_{i}^{(k+1)} + \cs_{i+1}^{(k+1)}\,, \\
  \cs_{i+2}^{(k+2)} &= \cs_{i+1}^{(k+1)} + \cs_{i+2}^{(k+1)}\,, \\
  \cs_j^{(k+2)} &= \cs_j^{(k+1)}\,, \qquad\qquad j \ne i \text{ or } i+2.
\end{align*}
We can rewrite the net result of~$\Kl_i\Kl_{i+1}$ on columns~$i$,
$i+1$ as
\begin{align*}
  \cs_{i}^{(k+2)} &= \cs_{i+1}^{(k+1)} + \cs_{i}^{(k)}\,, \\
  \cs_{i+1}^{(k+2)} &= \cs_{i}^{(k+1)} + \cs_{i+1}^{(k)}\,,
\end{align*}
where we used~$\cs_{i}^{(k+1)}=\cs_{i}^{(k)}$.  Now if we define the
sequence~$\as^{(k)}=\cs_{i}^{(k)}$ for~$k$ even,
and~$\as^{(k)}=\cs_{i+1}^{(k)}$ for~$k$ odd,
then~$\as^{(k+2)}=\as^{(k+1)}+\as^{(k)}$, the Fibonacci sequence.  If
the initial matrix~$M^{(0)}$ is the identity,
then~$\as^{(k)}=\fib_{k+2}$, where~$\fib_k$ is the~$k$th Fibonacci
number.  This is the largest of the column-sums,
so~$\hat\rho_k(\Klset,\lVert\cdot\rVert_1)=\fib_{k+2}$.  Inserting a
product by any other matrix but~$\Kl_i$,~$\Kl_{i+1}$ cannot cause the
largest column-sum to increase faster, since it will involve the sum
of elements that have not been added together.  (To complete the
proof we also need to show that repeated products such as~$\Kl_i^2$
have a smaller specral radius than~$\Kl_i\Kl_{i+1}$, which is
straightforward.)  Hence, we
have~$\hat\rho_k(\Klset,\lVert\cdot\rVert_1)=\fib_{k+2}$, so
that~$\JSR(\Klset)=\limsup_{k\rightarrow\infty}\fib_{k+2}^{1/k}=\phi_1$,
the golden ratio.  Since we have an explicit braid realising this TEPG
for~$\nn=3$ and~$4$, we conclude that the optimal TEPG is equal to the
golden ratio.

\begin{figure}
\begin{center}
  \includegraphics[width=.5\textwidth]{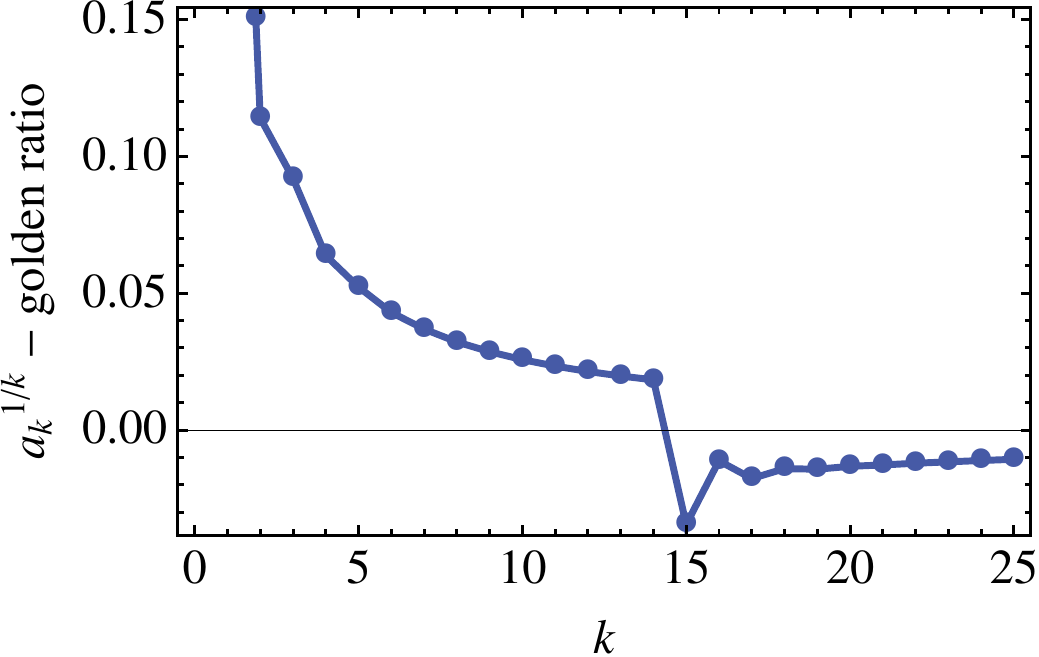}
\end{center}
\caption{Convergence of the sequence~$\as_k^{1/k}$ to the golden
  ratio: here~$\as_{k+2}=\as_{k+1}+\as_{k}$, except
  for~$\as_{15}=\as_{14}$.  The sequence approaches the golden ratio
  from below after the skipped term.}
\label{fig:conv_to_golden}
\end{figure}

For~$\nn>4$, we can approach golden ratio TEPG as close as we want by
using very long braids (see Section~\ref{sec:TEPG}); however, we have
not yet shown that there are \emph{no} braids realising a golden ratio
TEPG for~$\nn>4$.  To show this, observe that the earlier
sequence~$\fib_{k+2}^{1/k}$ approaches the golden ratio from above,
which is needed for an actual golden ratio TEPG to exist for
\emph{finite}~$k$, since~$\JSR(\Klset) \le
\l(\hat\rho_k(\Klset,\lVert\cdot\rVert_1)\r)^{1/k}$ for any~$k$.  If
we add the constraint of using all the generators, which is a
necessary condition for irreducibility (and hence for the mapping to
be pseudo-Anosov), then we `delay' the sequence by at least one term,
and it now approaches the golden ratio from below (see
Fig.~\ref{fig:conv_to_golden}).  Hence, for finite~$k$ the JSR can
never be equal to the golden ratio.  This means that for~$\nn>4$ we
have $\text{TEPG}<\phi_1$, strictly.

To find an upper bound on the topological entropy per operation (TEPO,
see Section~\ref{sec:TEPO}), we inflate the set~$\{\Kl_i\}$ into the
set~$\Llset=\{\Ll_j\}$ of all the operations that can be performed
simultaneously (i.e., products of~$\Kl_i$ that all commute, without
repeating any~$\Kl_i$).  For example, for~$\nn=6$ the set~$\Llset$
consists of~$12$ matrices:
\begin{equation*}
  \Llset = \Klset \cup
  \{\Kl_1\Kl_3,\Kl_1\Kl_4,\Kl_1\Kl_5,\Kl_2\Kl_4,\Kl_2\Kl_5,\Kl_3\Kl_5\}
  \cup \{\Kl_1\Kl_3\Kl_5\}
\end{equation*}
In general, the set~$\Llset$ has cardinality $\#\{\Ll_j\} =
\sum_{j=1}^\nn\binom{\nn-j}{j} = F_{\nn+1} - 1$, where~$F_{\nn}$ it
the~$\nn$th Fibonacci number.  Since the largest column-sum
in~$\Llset$ is always~$3$, we immediately get the bound~$\text{TEPO}
\le \log 3$ independent of~$\nn$.  If we use the 2-norm instead, we
get the result~$\text{TEPO} < \log(1+\sqrt{2})$ (the inequality in
this case is strict), again independent of~$\nn$.  Since the TEPO of
the sequence~\eqref{eq:maxTEPO} asymptotes to~$\log(1+\sqrt{2})$ for
large~$\nn$, then the optimal TEPO converges to~$\log(1+\sqrt{2})$ as
claimed in Section~\ref{sec:TEPO}.

Intuition suggests that the largest joint spectral radius in~$\Llset$
is given by the product~$\Llo\Lle$, with
\begin{equation*}
  \Llo = \Kl_1\Kl_3\Kl_5\cdots, \qquad
  \Lle = \Kl_2\Kl_4\Kl_6\cdots.
\end{equation*}
In fact the product~$\Llo\Lle$ gives exactly the same TEPO as the
sequence~\eqref{eq:maxTEPO} for each~$\nn$.  Hence, proving that the
JSR of the set~$\Ll$ is realised by the product~$\Llo\Lle$ for
all~$\nn$ is enough to prove the optimality of~\eqref{eq:maxTEPO}.
Proving that we have the JSR appears much harder than for the TEPG.

Note that it seems fortuitous that for both the optimal TEPG and TEPO
the upper bound given by the joint spectral radius of the incidence
matrices is sharp.  If this were not the case, then proving that we
have the TEPG or TEPO would involve more that just linear algebra
techniques.  Intuitively, one can argue that the braids realising the
optimal TEPG and TEPO avoid `cancellations' in growth of words
in~$\pi_1$, which is not surprising since such cancellations would
make the braid less efficient.

%\bibliographystyle{jlt}
%\bibliography{bib/journals_abbrev,bib/articles,silver}

\end{document}